\documentclass[11pt]{article}
\usepackage[T1]{fontenc}
\usepackage[utf8]{inputenc}
\usepackage{lmodern}
\usepackage{combelow}
\usepackage{amsmath,amssymb}
\usepackage{graphicx}
\usepackage[dvipsnames]{xcolor}
\usepackage{array,longtable,booktabs}
\usepackage{fancyvrb}
\usepackage[a4paper,margin=1in]{geometry}
\usepackage{enumitem}
\usepackage[hidelinks]{hyperref}
\usepackage{xurl}
\usepackage{seqsplit}
\usepackage{indentfirst}
\newcommand\runin[1]{\textit{#1}}
\newcommand\prooflabel[1]{\textit{#1}}
\newcommand\qedsym{\hfill\ensuremath{\square}}
\newcommand\thmskip{\par\medskip}

\newcommand\papertitle[3]{\begin{center}{\LARGE\bfseries #1\par}\vspace{1.2em}{\large #2\par}\vspace{0.6em}{#3\par}\end{center}\vspace{1em}}
\allowdisplaybreaks
\hypersetup{pdftitle={Global Convergence of Detailed Balanced Mass Action Systems}, pdfauthor={Haoran Lv, Chengjin Ni, Manqin Wang, Shenli He}, pdfkeywords={mass-action kinetics, detailed balance, persistence, Lyapunov function, global convergence, Lean formalization}}
\begin{document}
\papertitle{Global Convergence of Detailed Balanced Mass Action Systems}{Haoran Lv, Chengjin Ni, Manqin Wang, and Shenli He\par\vspace{0.4em}{\normalsize BitMiracle AI Lab}}{October 19, 2026}

\begin{abstract}
We give a self-contained proof that, for a finite detailed balanced mass-action system with positive rate constants, each positive stoichiometric compatibility class contains exactly one positive equilibrium and every positive trajectory converges to it, together with a Lean 4 formalization of the complete argument from persistence to convergence. Boundedness and persistence come from the barrier argument of OpenAI (2026) for weakly reversible systems: an arbitrary-tolerance affine approximation yields a finite-minimum barrier whose superlevel sets confine trajectories and give one logarithmic bound for all solutions from a given initial state. The Horn--Jackson dissipation identity identifies the positive equilibria as an exponential family with parameters in the orthogonal complement of the stoichiometric subspace, which gives uniqueness in each class. A bounded-speed descent argument replaces LaSalle's invariance principle; it uses neither uniqueness of solutions nor continuous dependence on initial data, and it places every cluster point in that family. The route given here is direct and elementary, and its main theorem is checked by the Lean kernel against explicit definitions. The proofs, the Lean development, and the manuscript were produced in ClawsGO Science, an agent harness that packages a complete research environment: Claude Opus 5.5 made the research decisions, GPT 6 Astra carried them out, and the Lean kernel checked every formal step.


\end{abstract}

\section{Introduction}

We consider finite mass-action reaction networks with positive constant rate coefficients and a positive detailed balanced equilibrium. The classical theory of Horn and Jackson \cite[pp.~81--116]{e11}, Feinberg \cite[pp.~187--194]{e5}, and Horn \cite[pp.~172--186]{e9} establishes the equilibrium structure of complex balanced systems, including existence and uniqueness in each positive compatibility class and local asymptotic stability. Horn \cite[pp.~125--137]{e10} formulated the global attractor conjecture, which asks whether every positive trajectory converges to the positive equilibrium in its compatibility class.

The entropy argument must be supplemented by an exclusion of boundary accumulation. Without this step, it permits an \(\omega\)-limit set contained in the positive equilibria \(\cup\) the boundary \cite[pp.~89--110]{e14}. Earlier partial results include the single linkage class case \cite{e1}, low-dimensional systems \cite{e13}, and strongly endotactic networks \cite{e8}. Craciun \cite[Section 8]{e3} gives a proof of the global attractor conjecture for complex balanced systems using toric differential inclusions.

Theorem 3.1 and Sections 3.1.1--3.1.7 build on OpenAI's work on boundedness and persistence of weakly reversible mass-action systems \cite{e15}; the detailed-balance argument developed here uses this theorem as input. The accompanying Lean modules for that part are adapted from the same work, with paths, namespaces, and principal declarations reorganized for this project.

\thmskip
\textbf{Theorem 1.1} (Main theorem). Assume that the network is finite, all rate constants are positive, and the mass-action system is detailed balanced at some \(x^{*} \in {\mathbb{R}}_{>0}^{d}\). Then, for every \(x_{0} \in {\mathbb{R}}_{>0}^{d}\), the following hold:

(i) The positive compatibility class \(\left( x_{0}+S \right) \cap {\mathbb{R}}_{>0}^{d}\) contains exactly one equilibrium \(\overline{x}\).

(ii) Equation (2.1) has a solution starting at \(x_{0}\) and defined on \(\lbrack 0,\infty)\).

(iii) Every such solution satisfies \(x(t) \rightarrow \overline{x}\) as \(t \rightarrow \infty\).

\runin{Contributions.} (1) A self-contained proof of Theorem 1.1 that joins persistence and the entropy argument into one chain. Theorem 3.1, stated for finite weakly reversible systems without detailed balance, supplies one compact container for all forward solutions from a given initial state (Corollary 3.15), and Steps 2--5 use Step 1 only through this container. The detailed-balance argument, this reduction, and the exact examples were developed for this paper.

(2) In Step 5, a bounded-speed descent argument for a non-increasing function bounded below replaces LaSalle's invariance principle. It uses neither continuous dependence on initial conditions nor uniqueness of solutions.

(3) A machine-checked proof of the whole chain from persistence to global convergence in Lean 4 \cite{e4} with Mathlib \cite{e12}. The formal main theorem quantifies over all global forward solutions, and the exact three-species example of Section 5.1 instantiates it. Section 4 gives the correspondence between the paper and the formal statements and the reproducible build.

\runin{Relation to previous work.} Existence, uniqueness, and the entropy structure are classical. Craciun's proof of the global attractor theorem \cite[Section 8]{e3} covers this case through toric differential inclusions; the route here is direct, uses elementary tools, and is machine-checked from persistence to convergence.

\runin{How the work was produced.} The work was carried out end to end in ClawsGO Science \cite{e16}, an agent harness that packages a complete research environment. Claude Opus 5.5 made the research decisions: it reduced Steps 2--5 to a single compact container for all solutions from a given initial state (Corollary 3.15), replaced LaSalle's invariance principle by the bounded-speed descent argument of Step 5, and chose to formalize the whole chain from persistence to convergence on top of the adapted persistence modules. GPT 6 Astra carried out these decisions: it wrote the proofs and the Lean modules, adapted the 71 persistence modules, ran the builds and axiom audits, and drafted the manuscript. The Lean kernel checked every formal step, and the human authors directed the work and reviewed the results.

\runin{Proof strategy and organization.} Section 2 fixes the notation. Section 3.1 constructs an affine barrier, proves confinement through feasible directions, and obtains a common logarithmic bound and a compact container for the trajectories. Section 3.2 proves entropy dissipation, parameterizes the positive equilibria, establishes uniqueness, and treats cluster points. Section 3.3 combines these steps. Section 4 records the formal statements, their correspondence with the paper, and the scope of verification. Section 5 gives examples and discusses persistence, complex balance, and boundary initial conditions.

\section{Mass Action Systems and Detailed Balance}

\subsection{Mass action systems}

Let there be \(d \geq 1\) species. A complex is a vector \(y \in {\mathbb{N}}^{d}\), and \(\mathcal{C} \subset{\mathbb{N}}^{d}\) is a finite set of complexes. A reaction network is a finite set \(\mathcal{R} \subset \mathcal{C}\times\mathcal{C}\). We write a reaction \(e = (y,y')\) as \(y \rightarrow y'\) and require \(y \neq y'\). Its source and target are \(s(e) = y\) and \(t(e) = y'\), respectively. Each reaction has a rate constant \(\kappa_{e} > 0\), and we put \(\kappa_{\mathrm{tot}} = \sum_{e\in \mathcal{R}}\kappa_{e}\). For \(x \in {\mathbb{R}}^{d}\), write the monomial \(x^{y} = \prod_{i = 1}^{d}x_{i}^{y_{i}}\) and the inner product \(a \cdot x = \sum_{i}a_{i}x_{i}\). The mass-action equation is
\[\dot{x} = f(x):= \sum_{y \rightarrow y' \in \mathcal{R}}\kappa_{y \rightarrow y'}\, x^{y}\,(y' - y).\tag{2.1}\]
The stoichiometric subspace is \(S = \operatorname{span}\left\{ y' - y:y \rightarrow y' \in \mathcal{R} \right\}\). For \(x_{0} \in {\mathbb{R}}_{>0}^{d}\), the set \(\left( x_{0}+S \right) \cap {\mathbb{R}}_{>0}^{d}\) is the positive stoichiometric compatibility class of \(x_{0}\). The map \(f\) is polynomial, and (2.1) gives \(f(x) \in S\) for every \(x\).

\thmskip
\textbf{Definition 2.1} (Solutions). Let \(x_{0} \in {\mathbb{R}}^{d}\).

(i) A map \(x:\mathbb{R}\rightarrow{\mathbb{R}}^{d}\) is a global forward solution starting at \(x_{0}\) if \(x(0) = x_{0}\) and, for every \(t \geq 0\), the map \(x\) is differentiable at \(t\) with derivative \(f\left( x(t) \right)\).

(ii) Let \(T \geq 0\). A map \(x\) is a solution on \(\lbrack 0,T\rbrack\) if \(x(0) = x_{0}\), the map \(x\) is continuous on \(\lbrack 0,T\rbrack\), and it is differentiable at every \(t \in (0,T)\) with derivative \(f\left( x(t) \right)\).

\thmskip
\textbf{Definition 2.1}(i) requires a two-sided derivative at \(t = 0\), so \(x\) must also be defined to the left of \(t = 0\). This is the formalization convention used in Section 4. Since \(f\) is polynomial, any classical forward solution extends a short distance backwards, so the convention does not restrict the result. The confinement argument of Lemma 3.10 and the descent argument of Step 5 apply to every solution without using uniqueness. The complete Lean proof also constructs a stronger forward-invariance property, including extension and agreement of finite solution segments; this construction uses standard ODE uniqueness for the locally Lipschitz polynomial vector field.

\subsection{Reversibility and detailed balance}

\thmskip
\textbf{Definition 2.2}.

(i) A network is reversible if \(y \rightarrow y' \in \mathcal{R}\) implies \(y' \rightarrow y\in \mathcal{R}\). In this case \(\overline{e} = (y',y)\) denotes the reverse of \(e = (y,y')\), and \(e \mapsto \overline{e}\) is an involution on \(\mathcal{R}\).

(ii) A network is weakly reversible if, for every \(y \rightarrow y' \in \mathcal{R}\), there is a directed path \(y' = c_{0} \rightarrow c_{1} \rightarrow \cdots \rightarrow c_{k} = y\) consisting of reactions in \(\mathcal{R}\), with \(k \geq 1\).

(iii) A system is detailed balanced if the network is reversible and there exists \(x^{*} \in {\mathbb{R}}_{>0}^{d}\) such that
\[\kappa_{e}\,\left( x^{*} \right)^{s(e)} = \kappa_{\overline{e}}\,\left( x^{*} \right)^{t(e)} =:K_{e}\quad\quad\left( e\in \mathcal{R} \right).\tag{2.2}\]
By definition, \(K_{e} = K_{\overline{e}} > 0\).

\thmskip
\textbf{Lemma 2.3}. Every reversible network is weakly reversible.

\prooflabel{Proof.} For \(e = (y,y')\), the reverse reaction \(\overline{e} = (y',y)\in \mathcal{R}\) itself gives a path from \(y'\) to \(y\) of length \(1\). \qedsym{}

\thmskip
\textbf{Remark 2.4} (Existence of detailed balance). Assume the network is reversible. Choose a representative direction \(\nu_{p} = y'_{p} - y_{p}\) for each reverse pair and put \(\rho_{p} = \ln\left( \kappa_{y_{p} \rightarrow y'_{p}}/\kappa_{y'_{p} \rightarrow y_{p}} \right)\). Setting \(u = \ln x^{*}\) componentwise, (2.2) is equivalent to the linear system \(\nu_{p} \cdot u = \rho_{p}\). This system is solvable if and only if every real linear relation \(\sum_{p}\alpha_{p}\nu_{p} = 0\), with \(\alpha_p\in\mathbb{R}\), implies \(\sum_{p}\alpha_{p}\rho_{p} = 0\). These are the Wegscheider conditions \cite[pp.~1819--1827]{e6}. In particular, if the vectors \(\nu_{p}\) are linearly independent, all positive choices of rate constants admit detailed balance.

\subsection{Notation}

We use \(| \cdot |\) for the maximum norm on \({\mathbb{R}}^{d}\). For \(h \in (0,1)\) and \(p \in {\mathbb{R}}^{d}\), write \(h^{p}:= \left( h^{p_{1}},\ldots,h^{p_{d}} \right) \in {\mathbb{R}}_{>0}^{d}\). The map \(p \mapsto h^{p}\) is a bijection from \({\mathbb{R}}^{d}\) to \({\mathbb{R}}_{>0}^{d}\) with inverse \(p_{i} = \ln x_{i}/\ln h\). It maps the cube \(\left\{ |p| \leq 1 \right\}\) to the box
\[B_{h}:= \left\lbrack h,h^{-1} \right\rbrack^{d},\]
and maps the boundary \(\left\{ |p| = 1 \right\}\) of the cube to the topological boundary \(\partial B_{h}\) of the box. The monomial definition gives
\[\left( h^{p} \right)^{z}/\left( h^{p} \right)^{y} = h^{\, p \cdot \left( z-y \right)}\quad\quad\left( y,z \in {\mathbb{N}}^{d} \right).\tag{2.3}\]

\section{Proof of the Main Theorem}

We prove Theorem 1.1 in five steps. Section 3.1 proves boundedness and persistence for weakly reversible systems (Theorem 3.1), without (2.2), and applies the result to detailed balanced systems to obtain a common compact container for all solutions from the fixed initial state (Corollary 3.15). Section 3.2 supplies Steps 2--5 of the Horn--Jackson argument: entropy dissipation, identification of positive equilibria, uniqueness in each class, and identification of cluster points. Section 3.3 finishes the proof. Steps 2--5 use Step 1 only through Corollary 3.15.

\subsection{Boundedness and persistence for weakly reversible systems}

\subsubsection{Statement}

\thmskip
\textbf{Theorem 3.1} (Logarithmic confinement). Assume \(d \geq 1\), the network is finite and weakly reversible, the rate constants satisfy \(\kappa_{e} > 0\), and \(x_{0} \in {\mathbb{R}}_{>0}^{d}\). Then:

(a) There exists a global forward solution starting at \(x_{0}\).

(b) There exists \(L\geq0\) such that every global forward solution starting at \(x_{0}\) is positive at all nonnegative times and satisfies
\[\left|\ln x_i(t)\right|\leq L\quad\quad(t\geq0,\ i=1,\ldots,d).\tag{3.1}\]
The same \(L\) works for all these solutions and all coordinates; it may depend on the network, rate constants, and \(x_{0}\).

Here a global forward solution is understood as in Definition 2.1(i). Theorem 3.1 assumes only weak reversibility and positive constant rates; detailed balance (2.2) is not required.

Exponentiating (3.1) places every trajectory in the compact box \([e^{-L},e^L]^d\) inside the positive orthant. Conversely, on any nonempty compact subset of the positive orthant the finitely many continuous functions \(z\mapsto|\ln z_i|\) have a common finite maximum. Thus logarithmic confinement expresses both boundedness and separation from the boundary. We will construct one compact plateau containing all trajectories from \(x_0\) and obtain \(L\) from it.

\subsubsection{Strategy}

We construct a barrier as a finite minimum of affine functions. Let \(\ell_{j}(x) = r_{j} \cdot x + c_{j}\) for \(j = 1,\ldots,n\) and put
\[F(x) = \min_{1 \leq j \leq n}\ell_{j}(x).\]
An index \(j\) is active at \(x\) if \(\ell_{j}(x) = F(x)\), equivalently \(\ell_{j}(x) \leq \ell_{k}(x)\) for every \(k\). Every minimizer is active when there is a tie. We construct data with the following properties.

\begin{itemize}
\item
  There is a constant branch \(j_{0}\), with \(r_{j_{0}} = 0\) and \(\ell_{j_{0}} \equiv M\), that is active at \(x_{0}\). Thus \(F \leq M\), and the plateau \(P = \left\{ F = M \right\}\) contains \(x_{0}\).
\item
  At every \(z\in B_h\), the vector \(f(z)\) is a feasible direction of the superlevel set \(\{y:F(y)\geq F(z)\}\): a sufficiently short forward segment \(z+s f(z)\) stays in that set. The Lipschitz continuity of \(F\) then makes \(F\) non-decreasing along trajectories.
\item
  The constant branch is inactive on \(\partial B_{h}\), so the plateau does not meet the boundary of the box.
\end{itemize}

Consequently, \(F\left( x(t) \right)\) is bounded below by \(M\) and above by \(M\), and the trajectory remains on the plateau. The plateau is a compact convex polyhedron in \(\operatorname{int}B_{h}\), on which the logarithmic coordinates are uniformly bounded, giving (3.1).

The feasible-direction criterion of Lemma 3.8 reduces this property to \(r_j\cdot f(z)\geq0\) for every constraint tight at level \(F(z)\), namely every active branch. To obtain these inequalities, write \(z=h^p\). By (2.3), \(z^y=h^{p\cdot y}\). For small \(h\), complexes minimizing \(p\cdot y\) have dominant monomials in \(f\). If the ordering induced by \(r_j\) agrees with that induced by \(p\), the leading contributions to \(r_j\cdot f\) point upwards from the lowest levels (Lemma 3.2). The main difficulty is to make every active direction close enough to \(p\) while fixing all directions before choosing \(h\) (Proposition 3.5).

\subsubsection{Network lemmas}

\thmskip
\textbf{Lemma 3.2} (Cut-flux dominance).

Assume that the network is weakly reversible, \(x \in {\mathbb{R}}_{>0}^{d}\), \(r \in {\mathbb{R}}^{d}\), \(\delta \geq 0\), and the following conditions hold:

(O) For all \(y,z\in \mathcal{C}\), the inequality \(r \cdot y < r \cdot z\) implies \(x^{z} \leq \delta\, x^{y}\).

(Q) For each reaction \(e\), we have \(\delta\,\kappa_{\mathrm{tot}} < \kappa_{e}\).

Then \(r \cdot f(x) \geq 0\).

\prooflabel{Proof.} Define the height \(a(y) = r \cdot y\). For \(c\in\mathbb{R}\), let
\[U_{c} = \left\{ e:\ a\left( s(e) \right) \leq c<a\left( t(e) \right) \right\},\quad\quad D_{c} = \left\{ e:\ a\left( t(e) \right) \leq c<a\left( s(e) \right) \right\},\]
be the upward and downward reactions crossing height \(c\), respectively, and put \(Q(c) = \sum_{e \in U_{c}}\kappa_{e}x^{s(e)} - \sum_{e \in D_{c}}\kappa_{e}x^{s(e)}\).

(i) We have \(Q(c) \geq 0\). This is immediate if \(D_{c} = \varnothing\). Otherwise choose \(e \in D_{c}\). Weak reversibility gives a directed path from \(t(e)\) to \(s(e)\). Its initial height is \(\leq c\) and its final height is \(>c\), so some reaction \(g\) on the path satisfies \(a\left( s(g) \right) \leq c < a\left( t(g) \right)\), that is, \(g \in U_{c}\). For any \(e' \in D_{c}\), we have \(a\left( s(e') \right) > c \geq a\left( s(g) \right)\), and (O) gives \(x^{s(e')} \leq \delta x^{s(g)}\). By (Q) and \(x^{s(g)} > 0\),
\[\sum_{e' \in D_{c}}\kappa_{e'}x^{s(e')} \leq \delta\,\kappa_{\mathrm{tot}}\, x^{s(g)} < \kappa_{g}\, x^{s(g)} \leq \sum_{e'' \in U_{c}}\kappa_{e''}x^{s(e'')}.\]
(ii) Sum over cuts. Arrange the distinct values in \(\left\{ a(y):y \in \mathcal{C} \right\}\) as \(a_{1} < \ldots < a_{m}\). If a reaction \(e\) satisfies \(a\left( s(e) \right) = a_{\sigma} < a_{\tau} = a\left( t(e) \right)\), then \(a_{\tau} - a_{\sigma} = \sum_{k = \sigma}^{\tau-1}\left( a_{k+1}-a_{k} \right)\), and \(e \in U_{a_{k}}\) holds exactly for \(k = \sigma,\ldots,\tau - 1\). The downward case is analogous; equal-height reactions cross no \(a_{k}\). Hence
\[r \cdot f(x) = \sum_{e}\kappa_{e}x^{s(e)}\left( a\left( t(e) \right)-a\left( s(e) \right) \right) = \sum_{k = 1}^{m-1}\left( a_{k+1}-a_{k} \right)\, Q\left( a_{k} \right)\  \geq \ 0.\]
\qedsym{}

Alternatively, (ii) can be expressed as an integral of \(Q(c)\) with respect to \(c\), using \(\int_{a\left( s(e) \right)}^{a\left( t(e) \right)}dc = a\left( t(e) \right) - a\left( s(e) \right)\), without sorting the heights. The integral and finite cut-sum formulations are equivalent. Condition (Q) is imposed reaction by reaction, so the empty reaction set needs no separate argument. One upward reaction dominates the total downward flux; reverse reactions need not be paired. This is why weak reversibility suffices.

\thmskip
\textbf{Lemma 3.3} (Monomial ordering from an exponent gap). Let \(h \in (0,1\rbrack\), \(p \in {\mathbb{R}}^{d}\), \(H\in\mathbb{R}\), and \(y,z \in {\mathbb{N}}^{d}\). If \(p \cdot \left( z-y \right) \geq H\), then \(\left( h^{p} \right)^{z} \leq h^{H}\,\left( h^{p} \right)^{y}\).

\prooflabel{Proof.} By (2.3), \(\left( h^{p} \right)^{z} = h^{p \cdot \left( z-y \right)}\left( h^{p} \right)^{y}\). For \(h \leq 1\), the map \(s \mapsto h^{s}\) is non-increasing, so \(h^{p \cdot \left( z-y \right)} \leq h^{H}\). \qedsym{}

\thmskip
\textbf{Lemma 3.4} (Sign tolerance). Let \(\mathcal{D} \subset{\mathbb{R}}^{d}\) be finite.

(a) There exists a function \(E:{\mathbb{R}}^{d} \rightarrow \left( 0,\frac{1}{2} \right\rbrack\) such that, for any \(r,p \in {\mathbb{R}}^{d}\), if \(\left| p_{i}-r_{i} \right| < E(r)\) for all \(i\), then, for every \(v\in \mathcal{D}\) satisfying \(r \cdot v \neq 0\), the numbers \(p \cdot v\) and \(r \cdot v\) have the same sign and \(|p \cdot v| > |r \cdot v|/2\).

(b) For every finite set \(\mathcal{A} \subset{\mathbb{R}}^{d}\), there exists \(H \in (0,1\rbrack\) such that \(H \leq |r \cdot v|/2\) for all \(r\in \mathcal{A}\) and \(v\in \mathcal{D}\) with \(r \cdot v \neq 0\).

\prooflabel{Proof.} (a) Put \(L = 1 + \max\left( \left\{ 0 \right\} \cup \left\{ \sum_{i}\left| v_{i} \right|:v \in \mathcal{D} \right\} \right)\). For fixed \(r\), let \(m(r) = \min\left\{ |r \cdot v|:\ v\in \mathcal{D},\ r \cdot v \neq 0 \right\}\), taking \(m(r) = 1\) when the set is empty, and define \(E(r) = \min\left\{ \frac{1}{2},\ m(r)/(2L) \right\} > 0\). If \(\left| p_{i}-r_{i} \right| < E(r)\) for every \(i\), and \(v\in \mathcal{D}\) satisfies \(r \cdot v \neq 0\), then \(v \neq 0\) and
\[\left| p \cdot v-r \cdot v \right| \leq \max_{i}\left| p_{i}-r_{i} \right|\sum_{i}\left| v_{i} \right| < E(r)\, L \leq \frac{1}{2}m(r) \leq \frac{1}{2}|r \cdot v|,\]
Thus \(|p \cdot v| > \frac{1}{2}|r \cdot v| > 0\), and \(p \cdot v\) and \(r \cdot v\) have the same sign. (b) Take the minimum of \(1\) and all the finitely many positive numbers \(|r \cdot v|/2\), taking \(H = 1\) if this collection is empty. \qedsym{}

The definition of \(L\) includes zero in the finite set, so it gives \(L = 1\) when \(\mathcal{D} = \varnothing\).

The function \(E\) is chosen pointwise; no continuity is required.

\subsubsection{Affine approximation}

\thmskip
\textbf{Proposition 3.5} (Uniform affine approximation).

For every function \(E:{\mathbb{R}}^{d} \rightarrow (0,\infty)\), there exist \(n \geq 1\), directions \(r_{1},\ldots,r_{n} \in {\mathbb{R}}^{d}\), functions \(c_{1},\ldots,c_{n}:\mathbb{R}\rightarrow\mathbb{R}\), and \(h_{0} > 0\) such that the following holds. For every \(h \in \left( 0,h_{0} \right)\), every \(p \in {\mathbb{R}}^{d}\) satisfying \(|p| \leq 1\), and every index \(j\) active at \(h^{p}\) for the affine family \(\{ r_{k} \cdot x + c_{k}(h)\}_{k}\), we have
\[\left| p_{i}-r_{j,i} \right| < E\left( r_{j} \right)\quad\quad(i = 1,\ldots,d).\]
The directions \(r_{j}\) and intercept functions \(c_{j}\) are fixed before \(h\) is chosen; only the values \(c_{j}(h)\) vary with \(h\). The conclusion holds for every active index, including ties, as required by the tight-constraint criterion in Lemma 3.8. The proposition depends only on the dimension \(d\) and tolerance \(E\), not on the reaction network. Section 3.1.7 proves it by induction on dimension, first constructing local label families near a fixed minimum and then joining them along a chain of intervals.

\subsubsection{Assembly of the barrier}

\thmskip
\textbf{Proposition 3.6} (Barrier data). Assume weak reversibility, \(\kappa_{e} > 0\), and \(x_{0} \in {\mathbb{R}}_{>0}^{d}\). There exist \(h \in (0,1)\), \(n \geq 1\), affine functions \(\ell_{j}(x) = r_{j} \cdot x + c_{j}\) for \(j = 1,\ldots,n\), and an index \(j_{0}\) such that:

(B1) We have \(r_{j_{0}} = 0\), and \(j_{0}\) is active at \(x_{0}\).

(B2) We have \(x_{0}\in\operatorname{int}B_{h}\), that is, \(h < x_{0,i} < h^{-1}\) for all \(i\).

(B3) Put \(F=\min_j\ell_j\). For every \(z\in B_h\), there exists \(\delta_z>0\) such that
\[F(z+s f(z))\geq F(z)\quad\quad(0\leq s\leq\delta_z).\]
The step size may depend on \(z\); the condition is imposed throughout \(B_h\).

(B4) For every \(x \in \partial B_{h}\), the index \(j_{0}\) is inactive at \(x\).

\prooflabel{Proof.} Order of choices. Put \(\mathcal{D} = \left\{ e_{1},\ldots,e_{d} \right\} \cup \left\{ z - y:y,z \in \mathcal{C} \right\}\), where \(e_{i}\) are the coordinate vectors. Choose, in order, \(E\) from Lemma 3.4(a); the data \(n,r_{j},c_{j}( \cdot ),h_{0}\) from Proposition 3.5 for this \(E\); and \(H \in (0,1\rbrack\) from Lemma 3.4(b) for \(\mathcal{A} = \left\{ r_{1},\ldots,r_{n} \right\}\). Finally choose \(h \in \left( 0,\min\left\{ h_{0},1 \right\} \right)\) such that
\[\left| \frac{\ln x_{0,i}}{\ln h} \right| < H\quad(i = 1,\ldots,d),\quad\quad h^{H}\kappa_{\mathrm{tot}} < \kappa_{e}\quad\left( e \in \mathcal{R} \right).\tag{3.2}\]
This is possible: as \(h \downarrow 0\), we have \(\left| \ln h \right| \rightarrow \infty\) and \(h^{H} \rightarrow 0\). Put \(p_{i}^{0} = \ln x_{0,i}/\ln h\). Then \(x_{0} = h^{p^{0}}\) and \(\left| p_{i}^{0} \right| < H \leq 1\). Fix the intercepts \(c_{j}:= c_{j}(h)\). Since the directions are fixed before \(h\) is chosen, the gap \(H\) is independent of \(h\), allowing both inequalities in (3.2) to hold.

Key estimate. Suppose \(|p| \leq 1\), the index \(j\) is active at \(h^{p}\), and \(v\in \mathcal{D}\) satisfies \(r_{j} \cdot v \neq 0\). Proposition 3.5 gives \(\left| p_{i}-r_{j,i} \right| < E\left( r_{j} \right)\), and Lemma 3.4 yields
\[\operatorname{sign}(p \cdot v) = \operatorname{sign}\left( r_{j} \cdot v \right),\quad\quad|p \cdot v| > \left| r_{j} \cdot v \right|/2 \geq H.\tag{3.3}\]
For (B3), let \(x \in B_{h}\) and write \(x = h^{p}\) with \(|p| \leq 1\). Let \(j\) be active at \(x\). Apply Lemma 3.2 with \(\delta = h^{H}\). Condition (Q) is the second inequality in (3.2). For (O), if \(y,z\in \mathcal{C}\) and \(r_{j} \cdot y < r_{j} \cdot z\), take \(v = z - y\in \mathcal{D}\). Then \(r_{j} \cdot v > 0\), so (3.3) gives \(p \cdot \left( z-y \right) > 0\) and \(p \cdot \left( z-y \right) \geq H\). Lemma 3.3 gives \(x^{z} \leq h^{H}x^{y}\). Therefore \(r_{j} \cdot f(x) \geq 0\).

Apply the finite-constraint criterion of Lemma 3.8 with level \(m=F(x)\), base point \(x\), and direction \(f(x)\). Its tight constraints are precisely the active branches just considered. It gives a positive step on which every \(\ell_j\) stays at least \(F(x)\), proving (B3). The criterion is an elementary affine calculation independent of the barrier construction.

For (B1), a finite non-empty family always has an active index. Choose \(j_{0}\) active at \(x_{0}\). If \(r_{j_{0},i} \neq 0\), take \(v = e_{i}\). Equation (3.3), with \(p = p^{0}\), gives \(\left| p_{i}^{0} \right| \geq H\), contradicting \(\left| p_{i}^{0} \right| < H\). Thus \(r_{j_{0}} = 0\).

For (B2), the inequalities \(\left| p_{i}^{0} \right| < 1\) and \(0 < h < 1\) give \(h = h^{1} < h^{p_{i}^{0}} < h^{-1}\).

For (B4), let \(x \in \partial B_{h}\) and write \(x = h^{p}\). Then \(|p| \leq 1\), and \(\left| p_{i} \right| = 1\) for some coordinate. If \(j_{0}\) were active at \(x\), Proposition 3.5 would give \(\left| p_{i}-0 \right| < E(0) \leq \frac{1}{2}\), a contradiction. \qedsym{}

\subsubsection{The plateau and confinement}

Take the data from Proposition 3.6 and put \(M = c_{j_{0}}\), \(F = \min_{j}\ell_{j}\), and
\[P = \left\{ x \in {\mathbb{R}}^{d}:\ \ell_{j}(x) \geq M,\ j = 1,\ldots,n \right\}.\]

\thmskip
\textbf{Lemma 3.7} (Plateau geometry). (a) We have \(F \leq M\) everywhere, and \(x \in P\) if and only if \(F(x) = M\), equivalently \(j_{0}\) is active at \(x\). (b) The set \(P\) is an intersection of finitely many closed half-spaces, hence a closed convex polyhedron, and \(x_{0} \in P\). (c) We have \(P\subset\operatorname{int}B_{h}\); in particular, \(P\) is compact and contained in \({\mathbb{R}}_{>0}^{d}\).

\prooflabel{Proof.} (a) Since \(r_{j_{0}} = 0\), we have \(\ell_{j_{0}} \equiv M\) and hence \(F \leq M\). The equality \(F(x) = M\) is equivalent to \(\ell_{j}(x) \geq M = \ell_{j_{0}}(x)\) for every branch. (b) This follows from the definition and (B1). (c) Parts (a) and (B4) imply \(P \cap \partial B_{h} = \varnothing\). Thus \(P = \left( P \cap \operatorname{int}B_{h} \right) \cup \left( P \smallsetminus B_{h} \right)\) is a disjoint union of relatively open sets. The set \(P\) is convex and therefore connected. By (B2), the first set contains \(x_{0}\) and is non-empty, so the second is empty. Thus \(P\) is a closed subset of the compact box \(B_{h}\) and is compact. \qedsym{}

\thmskip
\textbf{Lemma 3.8} (Feasible directions under finite affine constraints). Let \(J\) be finite, let \(\ell_j(y)=r_j\cdot y+c_j\) for \(j\in J\), and fix a level \(m\in\mathbb R\). Put \(Q_m=\{y:\ell_j(y)\geq m\text{ for all }j\in J\}\). For \(z\in Q_m\), a vector \(v\) satisfies
\[z+s v\in Q_m\quad(0\leq s\leq\delta)\quad\text{for some }\delta>0\]
if and only if \(r_j\cdot v\geq0\) for every tight constraint \(j\), that is, every \(j\) with \(\ell_j(z)=m\).

\prooflabel{Proof.} Write \(a_j=\ell_j(z)-m\geq0\) and \(b_j=r_j\cdot v\). Then
\[\ell_j(z+s v)-m=a_j+s b_j.\]
Suppose the tight constraints have nonnegative slopes. For \(a_j=0\), set \(\delta_j=1\); the displayed expression is nonnegative for all \(s\geq0\). For \(a_j>0\), set \(\delta_j=a_j/(|b_j|+1)>0\). If \(0\leq s\leq\delta_j\), then
\[a_j+s b_j\geq a_j-s|b_j|\geq s\geq0.\]
Thus \(\delta=\min(\{1\}\cup\{\delta_j:j\in J\})>0\) works for every constraint, including when \(J\) is empty. Conversely, a feasible segment and a tight constraint give \(0\leq a_j+\delta b_j=\delta b_j\), hence \(b_j\geq0\). \qedsym{}

For the barrier family, taking \(m=F(z)\) identifies the tight constraints with all active branches at \(z\). At a plateau point, \(m=M\) and \(Q_m=P\). The finite minimum defining \(\delta\) accounts both for ties and for the slack of the other constraints.

\thmskip
\textbf{Lemma 3.9} (From feasible directions to monotonicity). Let \(F=\min_{1\leq j\leq n}\ell_j\), where \(n\geq1\) and the \(\ell_j\) are affine. Suppose \(x:[a,b]\to\mathbb R^d\) is continuous and differentiable on \((a,b)\). If, for each \(t\in(a,b)\), the direction \(\dot x(t)\) is feasible in \(\{y:F(y)\geq F(x(t))\}\) at \(x(t)\), then \(F\circ x\) is non-decreasing on \([a,b]\).

\prooflabel{Proof.} The maximum norm gives
\[|F(y)-F(z)|\leq C|y-z|,\qquad C=\max_{1\leq j\leq n}\sum_i|r_{j,i}|.\]
Indeed, use a minimizing branch at \(z\) to bound \(F(y)-F(z)\), and then exchange \(y,z\). Put \(g=F\circ x\); it is continuous on \([a,b]\), because \(x\) is continuous and \(F\) is Lipschitz. For fixed \(t\in(a,b)\), differentiability gives \(x(t+s)=x(t)+s\dot x(t)+e_t(s)\), with \(|e_t(s)|/s\to0\) as \(s\downarrow0\). For every sufficiently small positive \(s\), feasibility and the Lipschitz estimate imply
\[g(t+s)\geq F(x(t)+s\dot x(t))-C|e_t(s)|\geq g(t)-C|e_t(s)|.\]
Consequently \(\liminf_{s\downarrow0}(g(t+s)-g(t))/s\geq0\). The feasible step may vary with \(t\); only the limit at this fixed \(t\) is used.

To pass from this local inequality to monotonicity, fix \(\eta>0\) and put \(g_\eta(t)=g(t)+\eta t\). If \(a<t_1<t_2<b\) and \(g_\eta(t_1)>g_\eta(t_2)\), the nonempty closed set \(\{s\in[t_1,t_2]:g_\eta(s)\geq g_\eta(t_1)\}\) has a largest point \(s^*<t_2\). At \(s^*\) the lower right Dini derivative is at least \(\eta>0\), so \(g_\eta(s)>g_\eta(s^*)\) for all sufficiently close \(s>s^*\). This contradicts maximality. Let \(\eta\downarrow0\) and use continuity at the endpoints. \qedsym{}

\thmskip
\textbf{Lemma 3.10} (Confinement). Let \(z \in P\) and \(T \geq 0\), and let \(x\) be a solution starting at \(z\) on \(\lbrack 0,T\rbrack\) in the sense of Definition 2.1(ii). Then \(x\left( \lbrack 0,T\rbrack \right) \subset P\).

\prooflabel{Proof.} The case \(T = 0\) is immediate. For \(T > 0\), put \(\tau = \sup\left\{ s \in \lbrack 0,T\rbrack:x\left( \lbrack 0,s\rbrack \right) \subset B_{h} \right\}\). By \(z \in P\subset\operatorname{int}B_{h}\) and continuity, \(\tau > 0\). Since \(B_{h}\) is closed, \(x\left( \lbrack 0,\tau\rbrack \right) \subset B_{h}\). For every \(t\in(0,\tau)\), condition (B3) at \(x(t)\) says that \(\dot x(t)=f(x(t))\) is feasible in the current superlevel set. Lemma 3.9 therefore gives \(F(x(t))\geq F(z)=M\) on \([0,\tau]\). Together with \(F\leq M\), this gives \(x(t)\in P\) throughout that interval. If \(\tau < T\), then \(x(\tau) \in P\subset\operatorname{int}B_{h}\) and continuity keeps \(x\) in \(B_{h}\) for a short interval after \(\tau\), contradicting the definition of \(\tau\). Hence \(\tau = T\). \qedsym{}

Confinement holds for every solution, with no uniqueness, global existence, or positivity assumed in advance. Positivity follows from \(B_{h} \subset {\mathbb{R}}_{>0}^{d}\).

\thmskip
\textbf{Lemma 3.11} (Global existence from compact confinement). Let \(K \subset {\mathbb{R}}^{d}\) be compact. Suppose that, for every \(z \in K\) and \(T \geq 0\), all solutions starting at \(z\) on \(\lbrack 0,T\rbrack\) remain in \(K\). Then a global forward solution exists from every \(z \in K\).

\prooflabel{Proof.} The polynomial map \(f\) is bounded and Lipschitz on a compact neighborhood of \(K\). The Picard--Lindelöf theorem gives a time \(\tau > 0\) independent of \(z \in K\) such that a \(C^{1}\) solution exists on \(\left\lbrack -\tau,\tau \right\rbrack\) from every \(z \in K\). Choose such a solution \(x^{(0)}\) from \(z\). By confinement, \(x^{(0)}(\tau) \in K\). Choose another local solution starting at \(x^{(0)}(\tau)\) and translate its forward segment to \(\lbrack\tau,2\tau\rbrack\), extending the forward solution to \(\lbrack 0,2\tau\rbrack\). Continue inductively. Adjacent pieces agree in value at their junction, and both derivatives equal \(f\) at that value, so the concatenation is differentiable there. This gives a curve on \(\left\lbrack -\tau,\infty \right)\) satisfying the equation at every \(t \geq 0\). Extend it arbitrarily to \(( - \infty, - \tau)\). \qedsym{}

\prooflabel{Proof of Theorem 3.1.} Take the data of Proposition 3.6 and the plateau \(P\). Lemmas 3.10 and 3.11 give a global forward solution from \(x_0\in P\), proving (a). For any global forward solution from \(x_0\), apply Lemma 3.10 on each finite interval to obtain \(x(t)\in P\) for all \(t\geq0\). By Lemma 3.7, \(P\) is nonempty, compact, and contained in the positive orthant. Each continuous function \(z\mapsto|\ln z_i|\) therefore attains a finite maximum on \(P\). Set
\[L=\max_{1\leq i\leq d}\ \max_{z\in P}|\ln z_i|\geq0.\]
The plateau was chosen from the network and \(x_0\) before choosing a solution. Thus this one constant bounds every trajectory, proving (b). \qedsym{}

\subsubsection{Proof of Proposition 3.5}

We first prove a sequential version on a general cube \(\lbrack a,b\rbrack^{m}\) (Lemma 3.12), and then obtain the uniform conclusion of Proposition 3.5 by compactness. We denote the dimension by \(m\) because the proof proceeds by induction on dimension.

\runin{Notation.} For an integer \(m \geq 1\), \(u \in {\mathbb{R}}^{m}\), and \(0 < h < 1\), continue to write \(h^{u} = \left( h^{u_{1}},\ldots,h^{u_{m}} \right)\). A label is a pair \(\lambda = \left( r_{\lambda},c_{\lambda} \right)\), where the slope \(r_{\lambda} \in {\mathbb{R}}^{m}\) is independent of \(h\) and the intercept \(c_{\lambda}(h)\) is a real function defined for all sufficiently small \(h > 0\). Given a finite non-empty label family \(\Lambda\), a fixed small \(h > 0\), and \(x \in {\mathbb{R}}^{m}\), we call \(\lambda \in \Lambda\) active at \(x\) if
\[r_{\lambda} \cdot x + c_{\lambda}(h) = \min_{\eta \in \Lambda}\left( r_{\eta} \cdot x+c_{\eta}(h) \right).\]
All minimizers are active, including ties. When finitely many label families are combined, restrict their intercepts to a common domain \(0 < h < h_{*}\). In every sequential statement below, discard finitely many initial terms so that all parameters lie in this common domain. Write \(\mu(u) = \min_{i}u_{i}\) and \(\mathbf{1} = (1,\ldots,1)\). The largest coordinate of \(h^{u}\) is \(h^{\mu(u)}\).

\thmskip
\textbf{Lemma 3.12} (Affine approximation with arbitrary tolerance).

Let \(m \geq 1\), \(a < b\), and let \(E:\lbrack a,b\rbrack^{m} \rightarrow (0,\infty)\) be any function. There exists a finite non-empty label family \(\Lambda\) with slopes \(r_{\lambda} \in \lbrack a,b\rbrack^{m}\) and intercepts satisfying \(c_{\lambda}(h) = o\left( h^{a} \right)\) as \(h \rightarrow 0^{+}\), with the following property. For every \(h_{\nu} \rightarrow 0^{+}\) and every sequence \(p^{\nu} \in {\mathbb{R}}^{m}\) converging to \(p \in \lbrack a,b\rbrack^{m}\), if a fixed \(\lambda \in \Lambda\) is active at \(h_{\nu}^{\, p^{\nu}}\) for every \(\nu\), then
\[\left| p-r_{\lambda} \right| < E\left( r_{\lambda} \right).\]
No regularity is assumed for \(E\), and the approximating sequence \(p^{\nu}\) need not lie in \(\lbrack a,b\rbrack^{m}\). This is necessary for induction: after projection, the sequence may approach a face of a smaller cube from either side. The conclusion holds for every label that remains active along the sequence, including tied labels.

\prooflabel{Proof.} We use strong induction on \(m\), invoking the induction hypothesis only in positive dimensions below \(m\). First fix a value \(t\) of the minimum exponent \(\mu(p)\) and construct a label family. Negative intercepts compensate for increasing slope components in coordinates where \(h^{p}\) is small. Compactness extends the guarantee to nearby values of \(\mu\) around \(t\). A finite interval construction then selects among these families. When \(m = 1\), Stage 1 uses only a singleton family, so the remaining stages also establish the induction base.

\runin{Stage 1. Construction at a fixed minimum.}

Fix \(t \in \lbrack a,b\rbrack\). We construct a finite family \(\Lambda_{t}\) whose slopes all satisfy \(\mu(r) = t\) and which contains the label \(\left( t\mathbf{1},0 \right)\). Its active labels satisfy the conclusion whenever the limit \(p \in \lbrack a,b\rbrack^{m}\) has \(\mu(p) = t\) and activity is taken within \(\Lambda_{t}\).

If \(m = 1\) or \(t = b\), take \(\Lambda_{t} = \left\{ \left( t\mathbf{1},0 \right) \right\}\). In either case, the only point of the cube with \(\mu(p) = t\) is \(t\mathbf{1}\), so the conclusion is immediate. Henceforth assume \(m > 1\) and \(t < b\).

A label has type \(k\) if exactly \(k\) slope components are strictly greater than \(t\) and all others equal \(t\). Type \(0\) consists only of \(\left( t\mathbf{1},0 \right)\). Construct the remaining types in the order \(k = 1,\ldots,m - 1\). After constructing type \(k - 1\), choose \(\beta_{k}\) such that
\[t < \beta_{k} < b,\quad\quad\beta_{k} < \beta_{k-1}\ \left( k>1 \right),\quad\quad\beta_{k} - t < E(r)\ \text{ for every type-}(k-1)\text{ slope } r.\]
There are only finitely many slopes of type \(k - 1\), each with a positive tolerance, so such a \(\beta_{k}\) exists. Only these finitely many tolerance values are compared; no positive lower bound for \(E\) on the entire cube is needed.

For each \(k\)-element subset \(U \subset \left\{ 1,\ldots,m \right\}\) and each \(\sigma \in \left\lbrack \beta_{k},b \right\rbrack^{U}\), let \(r(\sigma) \in \lbrack a,b\rbrack^{m}\) equal \(\sigma\) on \(U\) and \(t\) outside \(U\). Apply the induction hypothesis in dimension \(k < m\) to the cube \(\left\lbrack \beta_{k},b \right\rbrack^{U}\) and tolerance \(\sigma \mapsto E\left( r(\sigma) \right)\), ordering the coordinates of \(U\) naturally. For every resulting label \((\sigma,c')\) with \(c'(h) = o\left( h^{\beta_{k}} \right)\), include the label
\[r = r(\sigma),\quad\quad c(h) = - h^{\beta_{k}} + c'(h)\]
in type \(k\). Taking all coordinate subsets and all types gives the finite non-empty family \(\Lambda_{t}\). Since \(\beta_{k} > t\), every intercept is \(o\left( h^{t} \right)\). The negative term \(- h^{\beta_{k}}\) favors increasing the slope in coordinates of \(x\) that are small relative to \(h^{t}\). As \(\beta_{k}\) decreases, this preference becomes stronger at each successive type.

\runin{Stage 2. Verification at a fixed minimum.} Assume \(m > 1\), \(t < b\), \(h_{\nu} \rightarrow 0^{+}\), \(p^{\nu} \rightarrow p \in \lbrack a,b\rbrack^{m}\), and \(\mu(p) = t\). Fix a type \(k\) label \((r,c)\) that remains active in \(\Lambda_{t}\) along the sequence, and let \(U\) be its raised coordinate set. All asymptotic comparisons below are along this sequence; write \(h = h_{\nu}\). Whenever a limiting exponent lies strictly on one side of a threshold, convergence gives a fixed positive exponent gap for sufficiently large \(\nu\). We use only these gaps and do not replace \(h^{p_{i}^{\nu}}\) by \(h^{p_{i}}\), which convergence of exponents alone would not justify.

If \(k > 0\), comparison with \(\left( t\mathbf{1},0 \right)\) and activity give
\[\sum_{i \in U}\left( r_{i}-t \right)\, h^{p_{i}^{\nu}}\  \leq \ h^{\beta_{k}} - c'(h) = \left( 1+o(1) \right)\, h^{\beta_{k}}.\]
Every coefficient on the left is \(\geq \beta_{k} - t > 0\). If \(p_{i} < \beta_{k}\) for some \(i \in U\), a fixed \(\delta > 0\) satisfies \(p_{i}^{\nu} \leq \beta_{k} - \delta\) for all sufficiently large \(\nu\). Dividing by \(h^{\beta_{k}}\) then gives a term tending to infinity on the left, a contradiction. Therefore
\[p_{i} \geq \beta_{k}\quad\quad(i \in U).\]
Among labels constructed using the same \(U\), the contribution \(t\sum_{i \notin U}x_{i}\) from coordinates outside \(U\) and the intercept term \(- h^{\beta_{k}}\) are common. Thus \((\sigma,c')\) is active at \(h^{p_{U}^{\nu}}\) within the family supplied by induction. Its limiting exponent vector \(p_{U}\) lies in \(\left\lbrack \beta_{k},b \right\rbrack^{U}\). The induction hypothesis gives
\[\max_{i \in U}\left| p_{i}-r_{i} \right| < E(r).\]
Here it is essential to allow approximating sequences outside the cube: \(p_{U}^{\nu}\) need not lie in \(\left\lbrack \beta_{k},b \right\rbrack^{U}\), although its limit does. In particular, a limiting coordinate equal to \(\beta_{k}\) or \(b\) requires no strict inequality.

We next control coordinates outside \(U\), taking \(U = \varnothing\) for type \(0\). Suppose \(k < m - 1\). We claim that
\[p_{i} \leq \beta_{k+1}\quad\quad(i \notin U).\]
Otherwise there is some \(j \notin U\) with \(p_{j} > \beta_{k+1}\), so every limiting exponent on \(U \cup \left\{ j \right\}\) is strictly greater than \(\beta_{k+1}\). When \(k > 0\), use \(p_{i} \geq \beta_{k} > \beta_{k+1}\) for \(i \in U\); when \(k = 0\), no such check is needed. Choose any type \(k + 1\) label whose raised coordinates are \(U \cup \left\{ j \right\}\). Its affine value minus the value of \(\left( t\mathbf{1},0 \right)\) is
\[- h^{\beta_{k+1}} + o\left( h^{\beta_{k+1}} \right),\]
since \(h^{p_{i}^{\nu}} = o\left( h^{\beta_{k+1}} \right)\) in every raised coordinate. For the purported active label, the corresponding difference is zero when \(k = 0\) and is bounded below by \(- h^{\beta_{k}} + o\left( h^{\beta_{k}} \right)\) when \(k > 0\). In the latter case, \(h^{\beta_{k}} = o\left( h^{\beta_{k+1}} \right)\). Thus, for sufficiently large \(\nu\), the competitor has strictly smaller value, contradicting activity. This proves the claim. Outside \(U\) we have \(p_{i} \geq t\) and \(r_{i} = t\), so
\[\left| p_{i}-r_{i} \right| \leq \beta_{k+1} - t < E(r)\quad\quad(i \notin U),\]
where the strict inequality is precisely the condition imposed on \(\beta_{k+1}\) when constructing type \(k + 1\).

If \(k = m - 1\), only one coordinate lies outside \(U\). Each exponent on \(U\) is \(\geq \beta_{k} > t\), and \(\mu(p) = t\), so the remaining limiting exponent equals \(t\) and has zero error. Combining the cases gives \(\left| p-r \right| < E(r)\) for every type, including types \(0\) and \(m - 1\). Strictness is not needed at the threshold \(p_{i} = \beta_{k}\).

\runin{Stage 3. Extension to nearby minima.}

Fix \(t\) and the finite family \(\Lambda_{t}\). For each label \(\lambda = (r,c)\), let \(A_{\lambda} \subset \lbrack a,b\rbrack^{m}\) be the set of limiting points realized by sequences \(h_{\nu} \rightarrow 0^{+}\) and \(p^{\nu} \in {\mathbb{R}}^{m}\) such that \(\lambda\) is active in \(\Lambda_{t}\) at \(h_{\nu}^{\, p^{\nu}}\) and \(p^{\nu}\) converges to the given point. The set \(A_{\lambda}\) is closed. Indeed, if \(z_{n} \in A_{\lambda}\) and \(z_{n} \rightarrow z\), choose a point \(u_{n}\) and a parameter \(h_{n}\) from a realizing sequence for \(z_{n}\) such that
\[0 < h_{n} < 1/n,\quad\quad\left| u_{n}-z_{n} \right| < 1/n,\]
and \(\lambda\) is active at \(h_{n}^{\, u_{n}}\). Then \(u_{n} \rightarrow z\), so \(z \in A_{\lambda}\). This diagonal selection uses actual active points and does not require continuity of \(c(h)\).

Consequently, the set
\[B_{t} = \bigcup_{\lambda = (r,c) \in \Lambda_{t}}\left\{ p \in A_{\lambda}:\ \left| p-r \right| \geq E(r) \right\}\]
is compact. Every \(E(r)\) here is a fixed real number, so continuity of \(E\) is unnecessary. Stages 1 and 2 show that no point of \(B_{t}\) satisfies \(\mu(p) = t\). Continuity of \(\mu\) therefore gives \(\rho_{t} > 0\) such that no point of \(B_{t}\) satisfies
\[\mu(p) \in I_{t}:= \left( t-\rho_{t},\, t+\rho_{t} \right)\]
Indeed, if \(B_{t}\) is non-empty, choose \(\rho_{t}\) below the positive minimum of \(\left| \mu(p)-t \right|\) on \(B_{t}\). If \(B_{t}\) is empty, any positive radius works. Thus the limiting active-label conclusion holds whenever \(p \in \lbrack a,b\rbrack^{m}\) and \(\mu(p) \in I_{t}\).

If necessary, decrease \(\rho_{t}\) further so that every local intercept satisfies
\[c(h) = o\left( h^{\, t+\rho_{t}} \right).\tag{3.4}\]
For a non-singleton family, it suffices to impose \(t + \rho_{t} < \min_{1 \leq k \leq m-1}\beta_{k}\); singleton intercepts vanish. This stronger decay uses the constructed form \(c(h) = - h^{\beta_{k}} + o\left( h^{\beta_{k}} \right)\) and \(\beta_{k} > t\) and does not follow from \(c(h) = o\left( h^{t} \right)\) alone. Decreasing the radius preserves the preceding guarantee. We now have a correct local family near every possible value of \(\mu(p)\); it remains to combine finitely many families so that the appropriate one is selected.

\runin{Stage 4. Finite interval construction and gluing.}

Each local family gives the required approximation near its own minimum level. Add a common intercept shift to each family so that every label active in the combined family comes from an appropriate neighborhood. Overlap between adjacent neighborhoods allows equality at the transition cuts.

Choose a finite subcover of \(\lbrack a,b\rbrack\) from \(\{ I_{t}\}_{t \in \lbrack a,b\rbrack}\) and discard any selected interval contained in another, retaining one of any identical pair. Order the remaining centers as \(a \leq t_{1} < \ldots < t_{N} \leq b\) and write \(I_{t_{j}} = \left( L_{j},R_{j} \right)\). Non-containment forces the left and right endpoints to have the same order. Since \(t_{j} = \left( L_{j}+R_{j} \right)/2\), this is also the order of the centers. Hence
\[L_{1} < \ldots < L_{N},\quad\quad R_{1} < \ldots < R_{N}.\]
Adjacent intervals overlap. Otherwise \(R_{j} \leq L_{j+1}\), leaving points in \(\left\lbrack R_{j},L_{j+1} \right\rbrack \subset \left( t_{j},t_{j+1} \right)\) uncovered by every selected interval. The endpoint ordering, coverage, and the fact that all centers lie in \(\lbrack a,b\rbrack\) imply that the first interval contains \(a\) and the last contains \(b\).

For \(1 \leq j < N\), choose
\[q_{j} \in \left( \max\left\{ t_{j},L_{j+1} \right\},\ \min\left\{ t_{j+1},R_{j} \right\} \right).\]
This open interval is non-empty: the adjacent intervals overlap, each contains its center, and \(t_{j} < t_{j+1}\). Thus \(q_{j}\) belongs to both neighboring intervals and lies strictly between their centers. Put \(q_{0} = a\) and \(q_{N} = b\). Then
\[t_{j} < q_{j} < t_{j+1}\quad\left( 1 \leq j<N \right),\quad\quad\left\lbrack q_{j-1},q_{j} \right\rbrack \subset I_{t_{j}}\quad(1 \leq j \leq N).\tag{3.5}\]
When \(N = 1\), no cut is needed and the single interval contains all of \(\lbrack a,b\rbrack\).

Take the union of the selected families and, at level \(t_{j}\), replace each intercept \(c(h)\) by
\[\widetilde{c}(h) = c(h) - \sum_{\ell =1}^{j-1}h^{q_{\ell}}\tag{3.6}\]
The sum is empty for \(j = 1\). Every new intercept is \(o\left( h^{a} \right)\): use (3.4) and \(t_{j} + \rho_{t_{j}} > a\), while each shift exponent satisfies \(q_{\ell} > t_{\ell} \geq a\). Within one local family, a common shift does not change activity. Between adjacent families, the extra term \(- h^{q_{j}}\) selects the family on the appropriate side of the cut.

Suppose a fixed label \((r,c)\) from level \(t_{j}\) remains active in the combined family along \(h_{\nu} \rightarrow 0^{+}\) and \(p^{\nu} \rightarrow p \in \lbrack a,b\rbrack^{m}\). Let \(c\) be its original local intercept and continue to write \(h = h_{\nu}\). We claim that
\[q_{j-1} \leq \mu(p) \leq q_{j}.\]
If \(j < N\) and \(\mu(p) > q_{j}\), each \(h^{p_{i}^{\nu}}\) is \(o\left( h^{q_{j}} \right)\). By (3.4) and \(q_{j} \in I_{t_{j}}\), the local intercept is also \(o\left( h^{q_{j}} \right)\). Subtracting the value of the shifted label \(\left( t_{j+1}\mathbf{1},0 \right)\) gives
\[\left( r-t_{j+1}\mathbf{1} \right) \cdot h^{p^{\nu}} + c(h) + h^{q_{j}} = h^{q_{j}} + o\left( h^{q_{j}} \right) > 0\]
for sufficiently large \(\nu\), contradicting activity. If \(j > 1\) and \(\mu(p) < q_{j-1}\), subtract instead the shifted label \(\left( t_{j-1}\mathbf{1},0 \right)\). The difference is
\[\left( r-t_{j-1}\mathbf{1} \right) \cdot h^{p^{\nu}} + c(h) - h^{q_{j-1}}.\]
Every coefficient in the inner product is \(\geq t_{j} - t_{j-1} > 0\). There is a coordinate \(i\) with \(p_{i} < q_{j-1}\), so \(h^{p_{i}^{\nu}}/h^{q_{j-1}} \rightarrow \infty\). Also \(q_{j-1} < t_{j} < t_{j} + \rho_{t_{j}}\) gives \(c(h) = o\left( h^{q_{j-1}} \right)\). Dividing the difference by \(h^{q_{j-1}}\) therefore gives a quantity tending to \(+ \infty\), again contradicting activity. At \(j = 1\) or \(j = N\), the outer bounds follow from \(\mu(p) \in \lbrack a,b\rbrack\).

By (3.5), the claim implies \(\mu(p) \in I_{t_{j}}\), including equality at either cut. Labels at the same level receive the same shift, so activity in the combined family implies activity in the original local family. The Stage 3 guarantee for \(I_{t_{j}}\) gives \(\left| p-r \right| < E(r)\). Each contradiction uses a competitor with strictly smaller value, so the argument also applies to tied active labels. This completes the induction. \qedsym{}

\thmskip
\textbf{Corollary 3.13} (Uniform active-label bound).

Let \(m,a,b,E\) be as in Lemma 3.12 and fix its label family \(\Lambda\). There exists \(h_{0} > 0\) such that, for every \(0 < h < h_{0}\) and every \(p \in \lbrack a,b\rbrack^{m}\), each label active at \(h^{p}\) satisfies \(\left| p-r_{\lambda} \right| < E\left( r_{\lambda} \right)\). The family \(\Lambda\) is fixed before \(h_{0}\) is chosen.

\prooflabel{Proof.} Otherwise, for every \(\nu\) there exist \(0 < h_{\nu} < 1/\nu\), a point \(p^{\nu} \in \lbrack a,b\rbrack^{m}\), and a label \(\lambda_{\nu}\) active at \(h_{\nu}^{\, p^{\nu}}\) such that \(\left| p^{\nu}-r_{\lambda_{\nu}} \right| \geq E\left( r_{\lambda_{\nu}} \right)\). Since \(\Lambda\) is finite and \(\lbrack a,b\rbrack^{m}\) is compact, a subsequence has \(\lambda_{\nu} = \lambda\) fixed and \(p^{\nu} \rightarrow p \in \lbrack a,b\rbrack^{m}\). On this subsequence \(E\left( r_{\lambda} \right)\) is constant. Passing to the limit gives \(\left| p-r_{\lambda} \right| \geq E\left( r_{\lambda} \right)\), contrary to the strict inequality in Lemma 3.12. No regularity of \(E\) is used. \qedsym{}

\prooflabel{Proof of Proposition 3.5.}

Take \(m = d\), \(a = - 1\), and \(b = 1\). Apply Lemma 3.12 and Corollary 3.13 to the restriction of \(E\) to \(\left\lbrack -1,1 \right\rbrack^{d}\), obtaining a finite non-empty family \(\Lambda\) and \(h_{0} > 0\). Decrease \(h_{0}\) if needed so that all intercepts are defined on \(\left( 0,h_{0} \right)\), and extend them arbitrarily to the remaining values of \(h\), for example by \(0\); these values do not affect the conclusion. Enumerate \(\Lambda\) as \(\{\left( r_{j},c_{j} \right)\}_{j = 1}^{n}\) with \(n \geq 1\). The condition \(|p| \leq 1\) is equivalent to \(p \in \left\lbrack -1,1 \right\rbrack^{d}\), and the assertion that \(j\) is active at \(h^{p}\) for \(\{ r_{k} \cdot x + c_{k}(h)\}_{k}\) is exactly the activity defined above. Corollary 3.13 therefore gives \(\left| p_{i}-r_{j,i} \right| < E\left( r_{j} \right)\) for \(i = 1,\ldots,d\). \qedsym{}

\thmskip
\textbf{Proposition 3.5} enters Step 1 in two ways. The directions \(r_{j}\) are fixed before \(h\), so the gap \(H\) in Lemma 3.4(b) is independent of \(h\). The conclusion holds for every active index, so every tight constraint passes the feasible-direction test in Lemma 3.8. The tolerance \(E\) may be discontinuous, which allows Proposition 3.6 to impose sign conditions near finitely many hyperplanes simultaneously.

\subsubsection{A common compact container in the positive orthant}

We return to detailed balanced systems and fix \(x_{0} \in {\mathbb{R}}_{>0}^{d}\).

\thmskip
\textbf{Lemma 3.14} (Invariance of the compatibility class). Let \(x\) be a global forward solution starting at \(x_{0}\). Then \(x(t) - x_{0} \in S\) for every \(t \geq 0\).

\prooflabel{Proof.} Let \(\varphi\) be any linear functional satisfying \(\varphi|_{S} = 0\). Since \(f(x) \in S\), we have \(\frac{d}{ds}\varphi\left( x(s) \right) = \varphi\left( f\left( x(s) \right) \right) = 0\) for \(s \geq 0\). The mean value theorem gives \(\varphi\left( x(t) \right) = \varphi\left( x_{0} \right)\), that is, \(\varphi\left( x(t)-x_{0} \right) = 0\). In finite dimensions, a vector annihilated by every functional in the annihilator of \(S\) belongs to \(S\). \qedsym{}

This proof uses the mean value theorem and does not require integration.

\thmskip
\textbf{Corollary 3.15} (Step 1). Assume detailed balance and \(x_{0} \in {\mathbb{R}}_{>0}^{d}\). There exists \(L\geq0\) such that the set
\[K:=[e^{-L},e^L]^d\cap(x_0+S)\]
has the following properties. (a) A global forward solution exists from \(x_{0}\). (b) Every global forward solution from \(x_{0}\) satisfies \(x(t) \in K\) for \(t \geq 0\). (c) The set \(K\) is compact and \(K \subset {\mathbb{R}}_{>0}^{d}\). (d) There exists \(B > 0\) such that \(\left| f(x) \right| \leq B\) for every \(x \in K\).

\prooflabel{Proof.} Lemma 2.3 gives weak reversibility, so Theorem 3.1 supplies \(L\) and (a). Exponentiating (3.1) gives \(e^{-L}\leq x_i(t)\leq e^L\) for every solution and nonnegative time. Lemma 3.14 places these states in \(x_0+S\), proving (b). The box is compact and \(x_0+S\) is closed, so their intersection is compact; every coordinate on it is at least \(e^{-L}>0\). This proves (c). Continuity of \(f\) on \(K\) gives a finite speed bound; increasing it to a positive number proves (d). \qedsym{}

The detailed-balanced convergence argument uses Theorem 3.1 only through Corollary 3.15. For the fixed initial state \(x_{0}\), the set \(K\) is a common compact container for all its forward solutions. The invariant plateau \(P\) constructed earlier provides the stronger trapping property. Compact containment is sufficient for the subsequent convergence proof.

\subsection{The Horn Jackson argument}

Throughout this subsection and Section 3.3, assume detailed balance, with \(x^{*}\) and \(K_{e}\) as in (2.2). For \(x \in {\mathbb{R}}_{>0}^{d}\) and \(y \in {\mathbb{N}}^{d}\), define
\[L_{y}(x) = \sum_{i = 1}^{d}y_{i}\left( \ln x_{i} - \ln x_{i}^{*} \right) = \ln\frac{x^{y}}{\left( x^{*} \right)^{y}}\]
Then \(x^{y} = \left( x^{*} \right)^{y}\, e^{L_{y}(x)}\).

\subsubsection{Entropy monotonicity}

Define
\[V(x) = \sum_{i = 1}^{d}\left( x_{i}\left( \ln x_{i} - \ln x_{i}^{*} \right)-x_{i}+x_{i}^{*} \right),\quad\quad D(x) = \sum_{i = 1}^{d}\left( \ln x_{i} - \ln x_{i}^{*} \right)f_{i}(x)\quad\quad\left( x \in {\mathbb{R}}_{>0}^{d} \right).\tag{3.7}\]
The quantity \(D = \nabla V \cdot f\) is the directional derivative of \(V\) along the vector field.

\thmskip
\textbf{Lemma 3.16} (Entropy properties). (a) We have \(V \geq 0\) on \({\mathbb{R}}_{>0}^{d}\). (b) If \(x\) is differentiable at \(t\) with derivative \(f\left( x(t) \right)\) and \(x(t) \in {\mathbb{R}}_{>0}^{d}\), then \(s \mapsto V\left( x(s) \right)\) is differentiable at \(t\) with derivative \(D\left( x(t) \right)\). (c) The function \(D\) is continuous on \({\mathbb{R}}_{>0}^{d}\).

\prooflabel{Proof.} (a) For \(a,b > 0\), the inequality \(\ln u \leq u - 1\) with \(u = b/a\) gives \(\ln b - \ln a \leq b/a - 1\). Multiplication by \(a\) gives \(a\left( \ln a - \ln b \right) - a + b \geq 0\); sum componentwise. (b) Apply the chain rule componentwise: \(\frac{d}{dt}\left\lbrack x_{i}\left( \ln x_{i} - \ln x_{i}^{*} \right)-x_{i} \right\rbrack = {\dot{x}}_{i}\left( \ln x_{i} - \ln x_{i}^{*} \right) + {\dot{x}}_{i} - {\dot{x}}_{i}\). (c) The function \(\ln\) is continuous on the positive reals, and \(f\) is polynomial. \qedsym{}

\thmskip
\textbf{Lemma 3.17} (Dissipation identity). For \(x \in {\mathbb{R}}_{>0}^{d}\), we have
\[2D(x) = \sum_{e\in \mathcal{R}}K_{e}\,\left( L_{t(e)}-L_{s(e)} \right)\left( e^{L_{s(e)}}-e^{L_{t(e)}} \right)\  \leq \ 0,\tag{3.8}\]
where \(L_{y} = L_{y}(x)\). Every summand is \(\leq 0\) and equals \(0\) if and only if \(L_{s(e)} = L_{t(e)}\).

\prooflabel{Proof.} Interchanging the sums gives \(D(x) = \sum_{e}\kappa_{e}x^{s(e)}\sum_{i}\left( t(e)_{i}-s(e)_{i} \right)\left( \ln x_{i} - \ln x_{i}^{*} \right) = \sum_{e}\kappa_{e}x^{s(e)}\left( L_{t(e)}-L_{s(e)} \right)\). By (2.2),
\[\kappa_{e}x^{s(e)} = K_{e}\, e^{L_{s(e)}},\quad\quad\kappa_{\overline{e}}\, x^{t(e)} = \kappa_{\overline{e}}\left( x^{*} \right)^{t(e)}e^{L_{t(e)}} = K_{e}\, e^{L_{t(e)}}.\]
The map \(e \mapsto \overline{e}\) is an involution on \(\mathcal{R}\), so \(\sum_{e}G(e) = \sum_{e}G\left( \overline{e} \right)\) for any \(G\). Hence \(2D = \sum_{e}\left( G(e)+G\left( \overline{e} \right) \right)\), where \(G(e) = \kappa_{e}x^{s(e)}\left( L_{t(e)}-L_{s(e)} \right)\). Substitution gives
\[G(e) + G\left( \overline{e} \right) = K_{e}e^{L_{s(e)}}\left( L_{t(e)}-L_{s(e)} \right) + K_{e}e^{L_{t(e)}}\left( L_{s(e)}-L_{t(e)} \right),\]
which is (3.8). The function \(\exp\) is strictly increasing, so \(\left( b-a \right)\left( e^{a}-e^{b} \right) \leq 0\), with equality if and only if \(a = b\). Also \(K_{e} > 0\). \qedsym{}

Lemmas 3.16 and 3.17 show that, along every positive solution, \(t \mapsto V\left( x(t) \right)\) is differentiable with derivative \(D\left( x(t) \right) \leq 0\). It is therefore non-increasing and bounded below by \(0\).

\subsubsection{Zero dissipation and positive equilibria}

\thmskip
\textbf{Lemma 3.18} (Exponential parameterization of positive equilibria). Assume positive rates and detailed balance at \(x^*\). Define \(\Phi(u)=(x_i^*e^{u_i})_{i=1}^d\). Then
\[\{x\in\mathbb R_{>0}^d:f(x)=0\}
 =\{x\in\mathbb R_{>0}^d:D(x)=0\}=\Phi(S^\bot).\]
The map \(\Phi:S^\bot\to\{x>0:f(x)=0\}\) is bijective, with inverse \(x\mapsto\ln x-\ln x^*\), where logarithms are componentwise.

\prooflabel{Proof.} First let \(u\in S^\bot\) and \(x=\Phi(u)\). Since every reaction vector belongs to \(S\),
\[L_{t(e)}(x)-L_{s(e)}(x)=u\cdot(t(e)-s(e))=0.\]
The flux identities of Lemma 3.17 give \(\kappa_e x^{s(e)}=K_e e^{L_{s(e)}}=K_e e^{L_{t(e)}}=\kappa_{\overline e}x^{t(e)}\). Reindexing (2.1) by the reverse-reaction involution therefore gives
\[2f(x)=\sum_e(\kappa_e x^{s(e)}-\kappa_{\overline e}x^{t(e)})(t(e)-s(e))=0.\]
Thus \(\Phi(S^\bot)\) consists of positive equilibria, and every positive equilibrium satisfies \(D(x)=\nabla V(x)\cdot f(x)=0\).

Conversely, let \(x>0\) satisfy \(D(x)=0\), and put \(u=\ln x-\ln x^*\). In (3.8) every summand is nonpositive. Since \(K_e>0\), a nonzero logarithmic gap would make the sum strictly negative. Hence \(u\cdot(t(e)-s(e))=L_{t(e)}-L_{s(e)}=0\) for every reaction. The reaction vectors span \(S\), so \(u\in S^\bot\), and exponentiation gives \(x=\Phi(u)\). This proves equality of the three sets. Finally, \(x=\Phi(v)\) forces \(v_i=\ln x_i-\ln x_i^*\) for every \(i\), proving uniqueness of the parameter. \qedsym{}

The parameterization describes all positive equilibria, including those outside the compatibility class of \(x^*\). If there are no reactions, \(S=\{0\}\) and \(\Phi(\mathbb R^d)=\mathbb R_{>0}^d\), as required.

\subsubsection{Uniqueness in each compatibility class}

\thmskip
\textbf{Lemma 3.19} (Uniqueness in each compatibility class). If \(a,b \in {\mathbb{R}}_{>0}^{d}\), \(f(a) = f(b) = 0\), and \(a - b \in S\), then \(a = b\).

\prooflabel{Proof.} By Lemma 3.18, write \(a=\Phi(u)\) and \(b=\Phi(v)\) with \(u,v\in S^\bot\). Since \(u-v\in S^\bot\) and \(a-b\in S\),
\[0=(u-v)\cdot(a-b)=\sum_i x_i^*(u_i-v_i)(e^{u_i}-e^{v_i}).\]
Every summand is nonnegative, and strict monotonicity of \(\exp\), together with \(x_i^*>0\), makes it zero only when \(u_i=v_i\). Therefore \(u=v\) and \(a=b\). \qedsym{}

\subsubsection{Cluster points and convergence}

A point \(z\) is a cluster point of \(x(t)\) as \(t \rightarrow \infty\) if there exists \(t_{n} \rightarrow \infty\) such that \(x\left( t_{n} \right) \rightarrow z\).

\thmskip
\textbf{Lemma 3.20} (Descent argument). Let \(x\) be a global forward solution with \(x(t)\in K\) for all \(t\geq0\), where \(K\subset\mathbb R_{>0}^d\) is compact. Fix \(B>0\) such that \(|f(w)|\leq B\) on \(K\). If \(z\in K\) is a cluster point, then \(D(z)=0\).

\prooflabel{Proof.} Put \(W(t) = V\left( x(t) \right)\). Step 2 and the mean value theorem show that \(W\) is non-increasing on \(\lbrack 0,\infty)\) and satisfies \(W \geq 0\). Let \(c = \inf_{t \geq 0}W(t)\).

Suppose \(D(z) \neq 0\). Lemma 3.17 gives \(D(z) < 0\). Put \(\eta = - D(z)/2 > 0\). Continuity of \(D\) at \(z\) gives \(r > 0\) such that \(\left| w-z \right| < r\) implies \(D(w) < - \eta\); decrease the radius so that this neighborhood lies in the positive orthant. Put \(\delta = r/(2B)\). Choose \(T \geq 0\) with \(W(T) < c + \eta\delta\), and then choose \(t \geq T\) with \(\left| x(t)-z \right| < r/2\), using that \(z\) is a cluster point. For \(s \in \left\lbrack t,t+\delta \right\rbrack\), the mean value inequality gives \(\left| x(s)-x(t) \right| \leq B\left( s-t \right) \leq B\delta = r/2\), hence \(\left| x(s)-z \right| < r\) and \(\frac{d}{ds}W(s) = D\left( x(s) \right) < - \eta\). A further application of the mean value theorem gives
\[W\left( t+\delta \right) \leq W(t) - \eta\delta \leq W(T) - \eta\delta < c,\]
contradicting the definition of \(c\) as the infimum. \qedsym{}

A standard alternative uses LaSalle's invariance principle: the \(\omega\)-limit set is forward invariant, \(V\) is constant on it, and therefore \(D = 0\). That formulation uses continuous dependence on initial conditions. Lemma 3.20 uses only a speed bound.

\subsection{Completion of the proof}

\thmskip
\textbf{Theorem 3.21} (Main theorem). Theorem 1.1 holds. In particular, the equilibrium \(\overline{x}\) in part (i) has the property that every \(z\) in \(\left( x_{0}+S \right) \cap {\mathbb{R}}_{>0}^{d}\) satisfying \(f(z) = 0\) equals \(\overline{x}\).

\prooflabel{Proof.} Take \(L,K,B\) from Corollary 3.15 and choose a global forward solution \(x^{(1)}\). Since it remains in the compact set \(K\), it has a cluster point \(\overline{x}\in K\). Lemma 3.20 gives \(D(\overline{x})=0\), so Lemma 3.18 places \(\overline{x}\) in \(\Phi(S^\bot)\), the set of positive equilibria. Also \(\overline{x}\in K\subset(x_0+S)\cap\mathbb R_{>0}^d\). If \(z\) is another equilibrium in this class, then \(z-\overline{x}\in S\), and Lemma 3.19 gives \(z=\overline{x}\). This proves (i); part (ii) is Corollary 3.15(a).

For (iii), let \(x\) be any global forward solution starting at \(x_{0}\). It remains in \(K\). Since \(K\) is closed, every cluster point lies in \(K\), so Lemma 3.20 applies. The same argument shows that every cluster point is an equilibrium in \(\left( x_{0}+S \right) \cap {\mathbb{R}}_{>0}^{d}\) and therefore equals \(\overline{x}\). A trajectory in a compact set with a unique cluster point converges to that point. Hence \(x(t) \rightarrow \overline{x}\). \qedsym{}

\thmskip
\textbf{Remark 3.22}. (1) The proof also yields existence of a positive equilibrium in every positive class. This classical conclusion can alternatively be obtained by Birch-type convex analysis \cite{e7,e11}. (2) Theorem 3.1 assumes \(d \geq 1\). For \(d = 0\), the space \({\mathbb{R}}^{0}\) has one point, so the conclusion of the main theorem is immediate.

\thmskip
\textbf{Corollary 3.23} (Asymptotic stability within the class). Put \(\Pi = \left( \overline{x}+S \right) \cap {\mathbb{R}}_{>0}^{d}\). Then \(\overline{x}\) is the strict global minimizer of \(V|_{\Pi}\) and is Lyapunov stable in \(\Pi\). Together with Theorem 3.21, this implies that \(\overline{x}\) is globally asymptotically stable in \(\Pi\).

\prooflabel{Proof.} If \(d=0\), the class is a singleton and both conclusions are immediate. Assume \(d\geq1\), and use the maximum norm throughout this proof. The function \(V\) is strictly convex on \({\mathbb{R}}_{>0}^{d}\), with Hessian \(\operatorname{diag}\left( 1/x_{i} \right)\). Lemma 3.18 gives \(\nabla V\left( \overline{x} \right) = \ln\overline{x} - \ln x^{*} \in S^{\bot}\). Thus, for \(x \in \Pi \smallsetminus \left\{ \overline{x} \right\}\), strict convexity gives \(V(x) > V\left( \overline{x} \right) + \nabla V\left( \overline{x} \right) \cdot \left( x - \overline{x} \right) = V\left( \overline{x} \right)\). For stability, fix any \(\eta > 0\) and choose \(0 < \rho < \min\left\{ \eta,\min_{i}{\overline{x}}_{i} \right\}\). The closed ball \(\overline{B}\left( \overline{x},\rho \right)\) lies in the positive orthant. Its sphere intersected with \(\Pi\) is compact, since it equals the intersection of that compact sphere with the closed affine space \(\overline{x}+S\). If this intersection is empty, take \(\delta = \rho/2\); continuity prevents a trajectory in \(\Pi\) starting within distance \(\delta\) from exiting this ball. Otherwise put \(m = \min\left\{ V(x) - V\left( \overline{x} \right):x \in \Pi,\  \parallel x - \overline{x} \parallel = \rho \right\} > 0\). This minimum exists and is positive by compactness and strict minimality. By continuity of \(V\), choose \(0 < \delta < \rho\) such that \(\parallel x - \overline{x} \parallel < \delta\) implies \(V(x) - V\left( \overline{x} \right) < m\). For any such initial state in \(\Pi\), Theorem 3.21 gives a global forward solution. Corollary 3.15(b), applied to this initial state, places every global forward solution \(x\) in a compact subset of the positive orthant, so \(V\circ x\) is non-increasing, and Lemma 3.14 keeps \(x\) in \(\Pi\). If \(\parallel x(t)-\overline{x}\parallel\geq\rho\) for some \(t\geq0\), then \(\parallel x(0)-\overline{x}\parallel<\delta<\rho\) and continuity give \(t_1>0\) with \(\parallel x(t_1)-\overline{x}\parallel=\rho\) and \(x(t_1)\in\Pi\). Hence \(V(x(t_1))-V(\overline{x})\geq m>V(x(0))-V(\overline{x})\), contradicting monotonicity. Consequently every such solution satisfies \(\parallel x(t) - \overline{x} \parallel < \rho < \eta\) for all \(t \geq 0\). This proves Lyapunov stability with the required quantifiers. \qedsym{}

\section{Lean Formalization}

\subsection{Scope}

Theorem 1.1 is formalized in Lean 4 with Mathlib, together with every numbered step of Section 3 from Theorem 3.1 to Theorem 3.21 and the exact example of Section 5.1. The development has about 9,300 lines. About 7,900 of them form the persistence part, adapted from OpenAI's formalization \cite{e15}, which supplies Theorem 3.1; the rest was written for this paper and contains the entropy argument, the convergence proof, and the examples. Section 4.2 gives the formal main theorem and the definitions it depends on, Section 4.3 the correspondence with the paper and the points where the formal proof departs from the text, and Section 4.4 the verification and the build.

\subsection{Formal statement of the main theorem}
\begin{Verbatim}[commandchars=\|\[\],fontsize=\small,xleftmargin=1em]
structure DBCondition (N : CRN d)
    (|ensuremath[|kappa] : RxnIdx N |ensuremath[|to] |ensuremath[|mathbb[R]]) (c : Fin d |ensuremath[|to] |ensuremath[|mathbb[R]]) : Prop where
  hRev : HasReverses N
  hPos : InOpenOrthant c
  hEq : |ensuremath[|forall] e : RxnIdx N,
    |ensuremath[|kappa] e * xPow c e.val.1 = |ensuremath[|kappa] (oppRxn hRev e) * xPow c e.val.2

theorem db_unique_equilibrium_attracts
    (N : CRN d) (|ensuremath[|kappa] : RxnIdx N |ensuremath[|to] |ensuremath[|mathbb[R]]) (h|ensuremath[|kappa] : |ensuremath[|forall] e, 0 < |ensuremath[|kappa] e)
    {c : Fin d |ensuremath[|to] |ensuremath[|mathbb[R]]} (hDB : DBCondition N |ensuremath[|kappa] c)
    (x0 : Fin d |ensuremath[|to] |ensuremath[|mathbb[R]]) (hx0 : InOpenOrthant x0) :
    |ensuremath[|exists] cstar : Fin d |ensuremath[|to] |ensuremath[|mathbb[R]],
      InOpenOrthant cstar |ensuremath[|wedge] rateField N |ensuremath[|kappa] cstar = 0 |ensuremath[|wedge]
      cstar - x0 |ensuremath[|in] reactionSpan N |ensuremath[|wedge]
      (|ensuremath[|forall] z, InOpenOrthant z |ensuremath[|to] rateField N |ensuremath[|kappa] z = 0 |ensuremath[|to]
        z - x0 |ensuremath[|in] reactionSpan N |ensuremath[|to] z = cstar) |ensuremath[|wedge]
      (|ensuremath[|exists] x, SolvesForward N |ensuremath[|kappa] x0 x) |ensuremath[|wedge]
      |ensuremath[|forall] x, SolvesForward N |ensuremath[|kappa] x0 x |ensuremath[|to]
        Tendsto x atTop (|ensuremath[|mathcal[N]] cstar)
\end{Verbatim}

The statement is taken from \texttt{GAC/\allowbreak{}Paper.\allowbreak{}lean}, with the proof omitted. The Lean kernel checks that the proof establishes exactly this statement. Whether the statement expresses Theorem 1.1 then depends only on the definitions it uses; together with the statement they occupy fewer than sixty lines on top of Mathlib. The names in the listing are transparent abbreviations of these definitions, which read as follows.

\begin{itemize}[leftmargin=1.5em,itemsep=2pt]
\item \texttt{N : CRN d} is a network as in Section 2.1: a finite set of complexes in \(\mathbb{N}^{d}\) and a finite set of reactions, each an ordered pair of distinct complexes of the network. \texttt{RxnIdx N} is the type of its reactions, so \(\kappa\) assigns a rate constant to each reaction, and \texttt{h\(\kappa\)} requires all of them to be positive.
\item \texttt{xPow c y} is the monomial \(c^{y}=\prod_{i}c_{i}^{y_{i}}\), and \texttt{rateField N \(\kappa\)} is the vector field \(f\) of (2.1), with the reaction vectors \(y'-y\) computed in \(\mathbb{R}^{d}\).
\item \texttt{InOpenOrthant x} means \(x\in\mathbb{R}_{>0}^{d}\), and \texttt{reactionSpan N} is the stoichiometric subspace \(S\).
\item \texttt{HasReverses N} is reversibility (Definition 2.2(i)), \texttt{oppRxn} maps \(e\) to \(\overline{e}\), and \texttt{DBCondition N \(\kappa\) c} is detailed balance (2.2), with \(c\) in the role of \(x^{*}\).
\item \texttt{SolvesForward N \(\kappa\) x0 x} is Definition 2.1(i): \(x:\mathbb{R}\to\mathbb{R}^{d}\), \(x(0)=x_{0}\), and \(x\) has derivative \(f(x(t))\) at every \(t\geq 0\). It assumes neither positivity nor convergence.
\end{itemize}

With these readings, the first three conjuncts of the conclusion say that \(c^{\star}\) is a positive equilibrium in \(x_{0}+S\), and the fourth that it is the only one; this is Theorem 1.1(i). The last two conjuncts are Theorem 1.1(ii) and (iii). The final conjunct quantifies over all global forward solutions, so it does not rely on uniqueness of solutions.

Two instances check that the hypotheses can be met and that the definitions take the expected values. For the network of Section 5.1, the development proves detailed balance at the stated rates, the conservation law \(x_{A}+2x_{B}+3x_{C}\), and convergence to \(\left(\frac{1}{2},\frac{1}{2},\frac{1}{2}\right)\) from every positive initial state in the class \(x_{A}+2x_{B}+3x_{C}=3\), including the five initial states of Figure 1 written as exact rationals. A second instance treats \(A\rightleftharpoons B\).

\subsection{Correspondence with the paper}

Table 1 lists, for each numbered result, the Lean declarations that state it; names are relative to the namespaces \texttt{GAC.Paper}, \texttt{GAC.Persistence}, and \texttt{GAC.Reformulated}, and to \texttt{GAC.PaperExample} for the example rows. File locations, and the declarations behind individual proof steps, are recorded in \texttt{docs/\allowbreak{}PAPER\_\allowbreak{}MAPPING.\allowbreak{}md} in the repository. The formal proof follows Section 3, with the following departures.

\begin{itemize}[leftmargin=1.5em,itemsep=2pt]
\item Lemmas 3.9--3.10. Lean derives confinement from the monotonicity of a finite minimum along curves on which every active branch has nonnegative slope. The Lipschitz estimate and the Dini argument of Lemma 3.9 have no separate Lean counterpart; the conclusion is the same.
\item Proposition 3.6 (B3). Lean proves that every active branch has nonnegative slope along \(f\) at each point of \(B_{h}\). The feasible-segment form follows from Lemma 3.8, as in the text; Lean states it on the plateau.
\item Theorem 3.1. Lean states positivity of solutions separately from the logarithmic bound, because under the Mathlib convention \(\log 0=0\) the bound alone does not exclude zero coordinates. The forward-invariance construction in its proof uses ODE uniqueness, as noted after Definition 2.1.
\item Corollary 3.15. The Lean container is \([\varepsilon,\varepsilon^{-1}]^{d}\cap(x_{0}+S)\) for some \(0<\varepsilon<1\), rather than \([e^{-L},e^{L}]^{d}\cap(x_{0}+S)\), with the same conclusions. The case \(d=0\) is treated separately.
\item Lemma 3.18. The equality of the three sets is assembled from several declarations rather than stated as one.
\end{itemize}

The following are not formalized: Corollary 3.23; the Wegscheider criterion of Remark 2.4 and the claim in Section 5.1 that the example is detailed balanced for all positive rates; the discussions in Sections 5.2--5.5; and the numerical computations behind Figure 1.

\begingroup\small
\begin{longtable}{@{}>{\raggedright\arraybackslash}p{0.14\linewidth}>{\raggedright\arraybackslash}p{0.82\linewidth}@{}}
\caption{Numbered results and the Lean declarations that state them.}\label{tab:corr}\\
\toprule
\textbf{Paper} & \textbf{Lean declarations}\\
\midrule\endfirsthead
\toprule
\textbf{Paper} & \textbf{Lean declarations}\\
\midrule\endhead
\bottomrule\endlastfoot
1.1 / 3.21 & \texttt{db\_\allowbreak{}unique\_\allowbreak{}equilibrium\_\allowbreak{}attracts}\\[2pt]
2.3 & \texttt{strongConn\_\allowbreak{}of\_\allowbreak{}hasReverses}\\[2pt]
3.1 & \texttt{wr\_\allowbreak{}trajectory\_\allowbreak{}log\_\allowbreak{}bound}\newline \texttt{bounded\_\allowbreak{}persistent\_\allowbreak{}of\_\allowbreak{}weaklyReversible}\\[2pt]
3.1 / 3.15 & \texttt{compact\_\allowbreak{}positive\_\allowbreak{}iff\_\allowbreak{}log\_\allowbreak{}bound}\newline \texttt{solution\_\allowbreak{}family\_\allowbreak{}log\_\allowbreak{}confinement}\\[2pt]
3.2 & \texttt{Cuts.\allowbreak{}cut\_\allowbreak{}flux\_\allowbreak{}nonneg\_\allowbreak{}of\_\allowbreak{}ordered\_\allowbreak{}sources}\newline \texttt{massAction\_\allowbreak{}points\_\allowbreak{}inward\_\allowbreak{}of\_\allowbreak{}order}\\[2pt]
3.3 & \texttt{monomial\_\allowbreak{}bound\_\allowbreak{}of\_\allowbreak{}exponent\_\allowbreak{}gap}\\[2pt]
3.4 & \texttt{exists\_\allowbreak{}sign\_\allowbreak{}margin\_\allowbreak{}function}\newline \texttt{exists\_\allowbreak{}uniform\_\allowbreak{}sign\_\allowbreak{}gap}\\[2pt]
3.5 & \texttt{uniform\_\allowbreak{}affine\_\allowbreak{}approximation}\newline \texttt{Affine.\allowbreak{}uniform\_\allowbreak{}data\_\allowbreak{}of\_\allowbreak{}full\_\allowbreak{}certificates}\\[2pt]
3.6 & \texttt{exists\_\allowbreak{}minBarrierSpec}\newline \texttt{Barrier.\allowbreak{}exists\_\allowbreak{}barrier\_\allowbreak{}data}\\[2pt]
3.7 & \texttt{Barrier.\allowbreak{}BarrierData.\allowbreak{}plateau\_\allowbreak{}subset\_\allowbreak{}interior\_\allowbreak{}box}\newline \texttt{.\allowbreak{}isCompact\_\allowbreak{}plateau}\\[2pt]
3.8 & \texttt{feasible\_\allowbreak{}direction\_\allowbreak{}iff\_\allowbreak{}tight}\\[2pt]
3.6 / 3.8 & Plateau special cases:\newline \texttt{plateau\_\allowbreak{}active\_\allowbreak{}iff\_\allowbreak{}tight}\newline \texttt{plateau\_\allowbreak{}feasible\_\allowbreak{}direction\_\allowbreak{}iff\_\allowbreak{}active}\newline \texttt{plateau\_\allowbreak{}massAction\_\allowbreak{}feasible}\\[2pt]
3.9--3.10 & \texttt{finite\_\allowbreak{}minimum\_\allowbreak{}monotone\_\allowbreak{}of\_\allowbreak{}active\_\allowbreak{}slopes}\newline \texttt{Confinement.\allowbreak{}forward\_\allowbreak{}solution\_\allowbreak{}trapped}\\[2pt]
3.11 & \texttt{global\_\allowbreak{}solution\_\allowbreak{}of\_\allowbreak{}compact\_\allowbreak{}trap}\newline \texttt{global\_\allowbreak{}massAction\_\allowbreak{}of\_\allowbreak{}compact\_\allowbreak{}trap}\\[2pt]
3.12 & \texttt{Affine.\allowbreak{}exists\_\allowbreak{}full\_\allowbreak{}certificate}\\[2pt]
3.13 & \texttt{Affine.\allowbreak{}FullCertificate.\allowbreak{}uniform\_\allowbreak{}activity\_\allowbreak{}bound}\\[2pt]
3.14 & \texttt{trajectory\_\allowbreak{}in\_\allowbreak{}coset}\\[2pt]
3.15 & \texttt{GAC.\allowbreak{}exists\_\allowbreak{}compact\_\allowbreak{}class\_\allowbreak{}container}\\[2pt]
3.16 & \texttt{entropyFn\_\allowbreak{}ge\_\allowbreak{}zero}\newline \texttt{entropyFn\_\allowbreak{}deriv}\newline \texttt{entropyRate\_\allowbreak{}continuous}\\[2pt]
3.17 & \texttt{GAC.\allowbreak{}dissip\_\allowbreak{}eq\_\allowbreak{}sum}\newline \texttt{entropyRate\_\allowbreak{}double}\newline \texttt{entropyRate\_\allowbreak{}le\_\allowbreak{}zero}\\[2pt]
3.18 & \texttt{logGap\_\allowbreak{}vanish\_\allowbreak{}of\_\allowbreak{}rate\_\allowbreak{}zero}\newline \texttt{rateField\_\allowbreak{}zero\_\allowbreak{}of\_\allowbreak{}logGap}\newline \texttt{entropyRate\_\allowbreak{}eq\_\allowbreak{}zero\_\allowbreak{}iff}\newline \texttt{positive\_\allowbreak{}equilibria\_\allowbreak{}eq\_\allowbreak{}exp\_\allowbreak{}normal}\newline \texttt{exponential\_\allowbreak{}parameter\_\allowbreak{}unique}\\[2pt]
3.19 & \texttt{equilibria\_\allowbreak{}unique\_\allowbreak{}of\_\allowbreak{}exp\_\allowbreak{}parameterization}\\[2pt]
3.20 & \texttt{entropyRate\_\allowbreak{}zero\_\allowbreak{}at\_\allowbreak{}accumulation}\\[2pt]
5.1 & \texttt{net\_\allowbreak{}detailed\_\allowbreak{}balanced}\newline \texttt{total\_\allowbreak{}conserved}\newline \texttt{tends\_\allowbreak{}to\_\allowbreak{}equilibrium\_\allowbreak{}of\_\allowbreak{}total\_\allowbreak{}eq\_\allowbreak{}three}\\[2pt]
Fig. 1 & \texttt{figure\_\allowbreak{}initial\_\allowbreak{}states\_\allowbreak{}positive}\newline \texttt{figure\_\allowbreak{}initial\_\allowbreak{}states\_\allowbreak{}total}\newline \texttt{figure\_\allowbreak{}initial\_\allowbreak{}states\_\allowbreak{}converge}\\[2pt]
\end{longtable}
\endgroup

\subsection{Verification and reproducibility}

The project is available at \url{https://github.com/BitMiracle-AI/detailed-balance-gac} (tag \texttt{v1.0}). It pins Lean 4.34.1 and Mathlib commit \texttt{\seqsplit{d13f23b723b8a846827a245b89c10fc7d3f11612}}. In the \texttt{lean/} directory, \texttt{lake exe cache get} followed by \texttt{lake build GAC} compiles the project from source in a few minutes. Then \texttt{lake env lean Audit.lean} and \texttt{lake env lean GAC/Reformulated/Audit.lean} print the axioms used by the main theorem, the example theorems, and the principal intermediate results. Each depends only on \texttt{propext}, \texttt{Classical.choice}, and \texttt{Quot.sound}, and none on \texttt{sorryAx}. A scan of the sources finds no \texttt{sorry}, \texttt{admit}, additional axioms, \texttt{unsafe} declarations, \texttt{native\_decide}, or \texttt{implemented\_by}. The repository README describes the module layout and gives the complete commands.

\section{Examples and Discussion}

\subsection{Numerical example}

Consider the three-species network \(A + B \rightleftharpoons C\) and \(2A \rightleftharpoons B\), with rate constants \(\kappa_{1} = 2,\kappa_{2} = 1\) for the first pair and \(\kappa_{3} = 1,\kappa_{4} = \frac{1}{2}\) for the second. The reaction vectors \(\left( -1, - 1,1 \right)\) and \(\left( -2,1,0 \right)\) are linearly independent, so Remark 2.4 gives detailed balance for any positive rate constants. The space \(S^{\bot}\) is spanned by \((1,2,3)\), giving the conserved quantity \(x_{A} + 2x_{B} + 3x_{C}\). Equilibria satisfy \(x_{B} = \left( \kappa_{3}/\kappa_{4} \right)x_{A}^{2}\) and \(x_{C} = \left( \kappa_{1}/\kappa_{2} \right)x_{A}x_{B}\). In the class \(x_{A} + 2x_{B} + 3x_{C} = 3\), the equilibrium is \(\overline{x} = \left( \frac{1}{2},\frac{1}{2},\frac{1}{2} \right)\). Substitution expresses the conserved quantity as a strictly increasing function of \(x_{A}\), confirming uniqueness.

Figure 1 shows five positive trajectories computed with LSODA using relative tolerance \(10^{-10}\) and absolute tolerance \(10^{-12}\) on \(0 \leq t \leq 8\). The initial states are (2.997, 0.001, 0.001/3), (0.001, 1.499, 0.001/3), (0.001, 0.22, 2.559/3), (1, 1/2, 1/3), and (0.3, 0.3, 0.7). All lie in the stated class. The trajectories approach the equilibrium, and the entropy evaluated relative to that equilibrium decreases. The computation illustrates the theorem.

\begin{figure}[t]
\centering
\includegraphics[width=\linewidth]{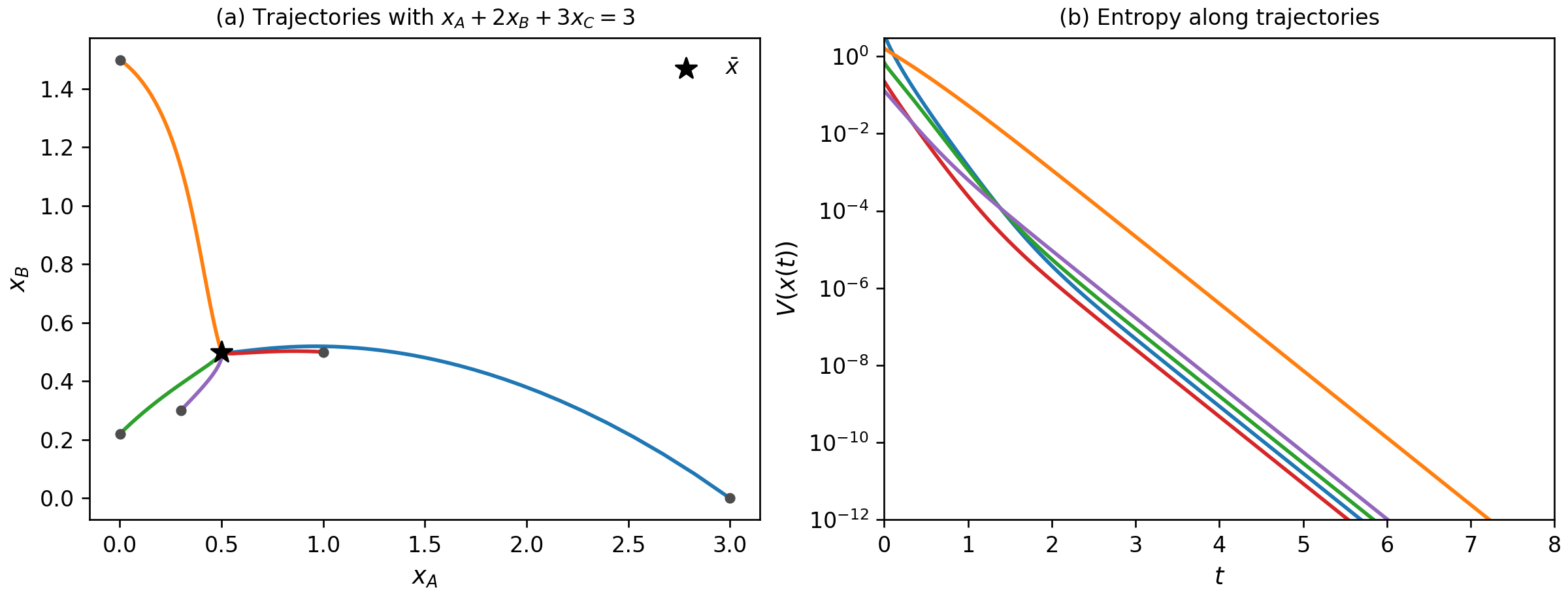}
\caption{Trajectories for \(A + B \rightleftharpoons C\) and \(2A \rightleftharpoons B\) in the class \(x_{A} + 2x_{B} + 3x_{C} = 3\). (a) Projection onto the \(\left( x_{A},x_{B} \right)\) plane; dots mark initial states and the star marks \(\overline{x} = \left( \frac{1}{2},\frac{1}{2},\frac{1}{2} \right)\). (b) A logarithmic plot of \(V\left( x(t) \right)\); the entropy panel is plotted down to \(10^{-12}\).}\label{fig:traj}
\end{figure}

\subsection{Why persistence is needed}

Without Step 1, the entropy argument permits an \(\omega\)-limit set contained in the positive equilibria \(\cup\) the boundary of the positive orthant \cite[pp.~89--110]{e14}. Boundary equilibria may exist, \(V\) is not differentiable there, and the magnitude of \(D\) can become arbitrarily small near the boundary.

\thmskip
\textbf{Example 5.1}. Consider \(2A \rightleftharpoons A + B\) with forward rate \(\kappa_{1}\) and reverse rate \(\kappa_{2}\). There is only one reaction vector, so detailed balance holds; take \(b^{*}/a^{*} = \kappa_{1}/\kappa_{2}\). The conserved quantity is \(a + b = T\) and the equation is \(\dot{a} = a\left( \kappa_{2}T-\left( \kappa_{1}+\kappa_{2} \right)a \right)\). Besides the positive equilibrium \(\left( \frac{\kappa_{2}T}{\kappa_{1} + \kappa_{2}},\frac{\kappa_{1}T}{\kappa_{1} + \kappa_{2}} \right)\), the closure of the class contains the boundary equilibrium \((0,T)\). By (3.8), as \(a \rightarrow 0\) we have \(D = O\left( a\left| \ln a \right| \right) \rightarrow 0\). Thus dissipation can approach zero at a boundary equilibrium as well as at a positive equilibrium. In this example, the logistic equation, or the specific shape of \(V\) on the class, excludes convergence to \((0,T)\) from positive initial states. Theorem 3.1 supplies the general compact-interior containment needed for the cluster-point argument.

\subsection{The role of persistence under detailed balance}

The function \(V\) extends continuously to the closed orthant under the convention \(0\ln 0 = 0\), and its sublevel sets there are compact. Along positive trajectories, \(V\left( x(t) \right) \leq V\left( x_{0} \right)\), so \(V\) alone gives boundedness in the detailed balanced case. What Step 5 needs from Step 1 is \(\liminf_{t \rightarrow \infty}x_{i}(t) > 0\) for \(i = 1,\ldots,d\). We use the full statement of Theorem 3.1 because it also provides global existence and matches the formal development.

\subsection{Complex balanced systems}

For a complex balanced system, there is a point \(x^{*} \in {\mathbb{R}}_{>0}^{d}\) at which total inflow equals total outflow at every complex. Replace the detailed-balance dissipation and equilibrium-identification steps as follows. Put \(w_{y} = x^{y}/\left( x^{*} \right)^{y}\). The inequality \(w\ln(w'/w) \leq w' - w\) and the complex balance equations give
\[D(x) = \sum_{e}\kappa_{e}\left( x^{*} \right)^{s(e)}w_{s(e)}\ln\frac{w_{t(e)}}{w_{s(e)}} \leq \sum_{e}\kappa_{e}\left( x^{*} \right)^{s(e)}\left( w_{t(e)}-w_{s(e)} \right) = 0,\]
Equality holds exactly when \(w_{s(e)} = w_{t(e)}\) for every reaction. Thus \(w_{y}\) is constant on each weakly connected component. Set \(a_{e} = \kappa_{e}\left( x^{*} \right)^{s(e)} > 0\). At zero dissipation, regrouping the vector field by complexes gives
\[f(x) = \sum_{y \in \mathcal{C}}y\left( \sum_{t(e) = y}a_{e}w_{s(e)} - \sum_{s(e) = y}a_{e}w_{y} \right) = \sum_{y \in \mathcal{C}}y\, w_{y}\left( \sum_{t(e) = y}a_{e} - \sum_{s(e) = y}a_{e} \right) = 0.\]
The first equality groups incoming and outgoing contributions at each complex. On an incoming edge, the equality condition gives \(w_{s(e)} = w_{y}\); complex balance makes each final bracket zero. Conversely, \(f(x) = 0\) implies \(D(x) = 0\). At every positive equilibrium, \(\ln\left( x/x^{*} \right) \cdot \left( t(e) - s(e) \right) = 0\) on each reaction, so the logarithmic uniqueness argument of Lemma 3.19 applies. Complex balance implies weak reversibility \cite[pp.~187--194]{e5}; \cite[pp.~172--186]{e9}, allowing Theorem 3.1 and the unchanged cluster-point argument to apply. This recovers the convergence conclusion covered by Craciun \cite[Section 8]{e3}.

\subsection{Boundary initial states}

The results concern strictly positive initial states. If a nonempty set of initially absent species is a siphon, meaning that every reaction producing a species in the set has a species of the set in its source complex \cite{e2}, uniqueness of solutions of the polynomial system keeps those species absent, and the trajectory stays on the boundary for as long as the solution exists. In Example 5.1, \(\left\{ A \right\}\) is a siphon. The global attractor conjecture likewise concerns positive initial states.

\subsection{Logical dependencies}

Section 3 presents the mathematical proof, with the geometric input isolated in Theorem 3.1. Section 4 records the formalization, its correspondence with the paper, and the reproducible build. The complete Lean proof uses standard ODE uniqueness in its strong-invariance construction, whereas the direct descent argument of Step 5 has no such dependency. Steps 2--5 apply to any global positive trajectory contained in a compact subset of its compatibility class. Suitable siphon criteria \cite{e2} or low-dimensional results \cite{e13} may establish the needed persistence for particular networks under their respective assumptions.

\subsection{Development in ClawsGO Science}

The harness closes the loop between proposing and checking. Each formal step written by GPT 6 Astra went directly to the Lean kernel, and the full build and axiom audits of Section 4.4 certify the result, including the adaptation of the 71 upstream modules under new paths, namespaces, and declaration names. For the handwritten arguments and the correspondence between the paper and the code, the harness ran three independent review sessions in parallel, covering Sections 2--3.1, Sections 3.2--5, and the correspondence with Lean; every finding is incorporated in the present version.

\section*{Code and data availability}
The Lean 4 project of Section 4 is publicly available at \url{https://github.com/BitMiracle-AI/detailed-balance-gac} under the Apache License 2.0; the version described in this paper is tagged \texttt{v1.0}, and Section 4.4 gives the commands that reproduce the check. The project and the numerical recomputation of Section 5.1, cross-checked with a second solver, were produced in ClawsGO Science \cite{e16}.

\section*{Acknowledgements and use of AI}
We acknowledge OpenAI's work on boundedness and persistence of weakly reversible mass-action systems \cite{e15}, from which Theorem 3.1 and Sections 3.1.1--3.1.7 derive. Seventy-one Lean modules of that work, at commit \texttt{\seqsplit{1581d4cbdd69a0faf34b2089614fe2a814316a6a}}, are adapted into \texttt{GAC/Persistence/}, with renamed paths, namespaces and principal declarations and with modification notices. The Apache-2.0 license is retained in \texttt{GAC/Persistence/LICENSE}; the old/new paths and separate upstream/current hashes are recorded in \texttt{lean/docs/SOURCE\_MAP.md} and its machine-readable companion.

The proofs, the Lean development, and the manuscript were produced in ClawsGO Science \cite{e16}, with Claude Opus 5.5 making the research decisions and GPT 6 Astra executing them. Together they developed the detailed-balance argument and its connection with Theorem 3.1, the logarithmic bounds, feasible directions, and exponential parameterization in the main text, wrote and audited the Lean 4 development (the \texttt{GAC} modules and the manuscript and coordinate/geometric interfaces), constructed the exact examples and the numerical illustration, and drafted and translated the manuscript. The human authors directed the work, checked the arguments and the correspondence between the Lean statements and the mathematical definitions, and take full responsibility for the content. No AI system is an author of this paper.

\end{document}